# Measurement Techniques for Thermal Conductivity and Interfacial Thermal Conductance of Bulk and Thin Film Materials

Dongliang Zhao, Xin Qian, Xiaokun Gu, Saad Ayub Jajja, Ronggui Yang*

*Department of Mechanical Engineering, University of Colorado, Boulder, CO 80309-0427*

*Corresponding author: ronggui.yang@colorado.edu

## Abstract

Thermal conductivity and interfacial thermal conductance play crucial roles in the design of engineering systems where temperature and thermal stress are of concerns. To date, a variety of measurement techniques are available for both bulk and thin film solid-state materials with a broad temperature range. For thermal characterization of bulk material, the steady-state method, transient hot-wire method, laser flash diffusivity method, and transient plane source method are most used. For thin film measurement, the 3ω method and the transient thermoreflectance technique including both time-domain and frequency-domain analysis are widely employed. This work reviews several most commonly used measurement techniques. In general, it is a very challenging task to determine thermal conductivity and interfacial thermal conductance with less than 5% error. Selecting a specific measurement technique to characterize thermal properties needs to be based on: 1) knowledge on the sample whose thermophysical properties is to be determined, including the sample geometry and size, and the material preparation method; 2) understanding of fundamentals and procedures of the testing technique, for example, some techniques are limited to samples with specific geometries and some are limited to a specific range of thermophysical properties; 3) understanding of the potential error sources which might affect the final results, for example, the convection and radiation heat losses.

*Keywords:* Bulk solid materials; Thin films; Thermal conductivity; Thermal contact resistance; Thermal boundary resistance; Interfacial thermal conductance

## Nomenclature

$a$ width, m

$A$ cross-sectional area, $\mathrm{m}^2$

$c_p$ specific heat capacity, $\mathrm{J\ kg^{-1}\ K^{-1}}$

$C$ volumetric heat capacity, $\mathrm{J\ m^{-3}\ K^{-1}}$

$d$ thickness, m

$G$ interfacial thermal conductance, $\mathrm{W\ m^{-2}\ K^{-1}}$

$G_{det}$ gain of photodetect

$H$ height, m

$I$ electric current, A

$I_0$ electric current amplitude, A

$k$ thermal conductivity, $\mathrm{W\ m^{-1}\ K^{-1}}$

$K$ thermal conductance, $\mathrm{W\ K^{-1}}$

$L$ length, m

$L_p$ thermal penetration depth, m

$p$ heating power, W

$P_1$ power of pump beam, W

$P_2$ power of probe beam, W

$\boldsymbol{q}$ heat flux, $\mathrm{W\ m^{-2}}$

$q$ complex thermal wave number, $\mathrm{m^{-1}}$

$Q$ heat transfer rate, W

$r$ radius, m

$R$ thermal resistance, $\mathrm{W\ K^{-1}}$

$R_e$ electrical resistance, Ω

$R_1$ reflectivity at the wavelength of pump beam

$R_2$ reflectivity at the wavelength of probe beam

$S_p$ sensitivity of signals to parameter $p$

$t$ time, s

$T$ temperature, °C or K

$T_h$ hot-side temperature, °C or K

$T_c$ cold-side temperature, °C or K

$T_\infty$ ambient temperature, °C or K

$V$ voltage, V

$V_1, V_2$ voltmeter, V

$V_{in}$ in-phase signal, mV

$V_{out}$ out-of-phase signal, mV

$w_0$ pump beam radius, μm

$w_1$ probe beam radius, μm

$W$ dimensionless number defined in Eq. (15)

***Greek symbols***

$\alpha$ thermal diffusivity, $\mathrm{m^2\ s^{-1}}$

$\alpha_R$ temperature coefficient of resistance, $\mathrm{K^{-1}}$

$\beta$ constant coefficient determined by Eq. (29)

$\delta$ index of direction in transient thermoreflectance method

$\Delta T$ temperature difference, °C or K

$\Delta T_0$ amplitude of temperature change, °C or K

$\Delta T_{pp}$ peak-to-peak amplitude of the temperature signal between two envelope curves in pulsed power technique, °C or K

$\eta$ dimensionless number defined in Eq. (16)

$\theta$ temperature after Fourier transform

$\kappa$ thermal conductivity, $\mathrm{W\ m^{-1}\ K^{-1}}$

$\lambda$ thermal conductivity, $\mathrm{W\ m^{-1}\ K^{-1}}$

$\tau$ half period of the heating current in pulsed power technique, s

$\tau_d$ the delay time in TDTR method, ns

$\phi$ defined as $\sqrt{t\alpha/r^2}$ in transient plane source method

$\omega$ frequency, Hz

$\omega_0$ modulation frequency, Hz

$\omega_s$ repetition frequency, Hz

***Subscript***

| | |
|---|---|
| $b$ | bottom |
| $f$ | associated with thin film |
| $loss$ | heat loss |
| $p$ | represent interface thermal conductance, heat capacity or thermal conductivity for sensitivity parameter |
| $r$ | coordinate |
| $s$ | standard |
| $S$ | associated with substrate |
| $S+f$ | associated with thin film on substrate |
| $t$ | top |
| $z$ | coordinate |
| $\parallel$ | in-plane |
| $\perp$ | cross-plane |

## 1. Introduction

Thermal conductivity (denoted as $k$, $\kappa$ or $\lambda$) measures the heat conducting capability of a material. As shown in Figure 1a, it can be defined as the thermal energy (heat) $Q$ transmitted through a length or thickness $L$, in the direction normal to a surface area $A$, under a steady-state temperature difference $T_h - T_c$. Thermal conductivity of a solid-phase material can span for several orders of magnitude, with a value of ~0.015 $\mathrm{W/m \cdot K}$ for aerogels at the low end to ~2000 $\mathrm{W/m{\cdot}K}$ for diamond and ~3000 $\mathrm{W/m{\cdot}K}$ for single-layer graphene at the high end, at room temperature. Thermal conductivity of a material is also temperature-dependent and can be directional-dependent (anisotropic). Interfacial thermal conductance (denoted as $K$ or $G$) is defined as the ratio of heat flux to temperature drop across the interface of two components. For bulk materials, the temperature drop across an interface is primarily due to the roughness of the surfaces because it is generally impossible to have "atomically smooth contact" at the interface as shown in Figure 1b. Interfacial thermal conductance of bulk materials is affected by several factors such as surface roughness, surface hardness, impurities and cleanness, the thermal conductivity of the mating solids and the contact pressure [1]. For thin films, the temperature drop across an interface can be attributed to the bonding strength and material difference. Note that thermal contact resistance and thermal boundary resistance (or Kapitza resistance [2]) are usually used to describe heat conduction capability of an interface in bulk materials and thin films, respectively. Interfacial thermal conductance is simply the inverse of thermal contact/boundary resistance. Knowledge of thermal conductivity and interfacial thermal conductance and their variation with temperature are critical for the design of thermal systems. In this paper, we review measurement techniques for characterizing thermal conductivity and interfacial thermal conductance of solid-state materials in both bulk and thin film forms.

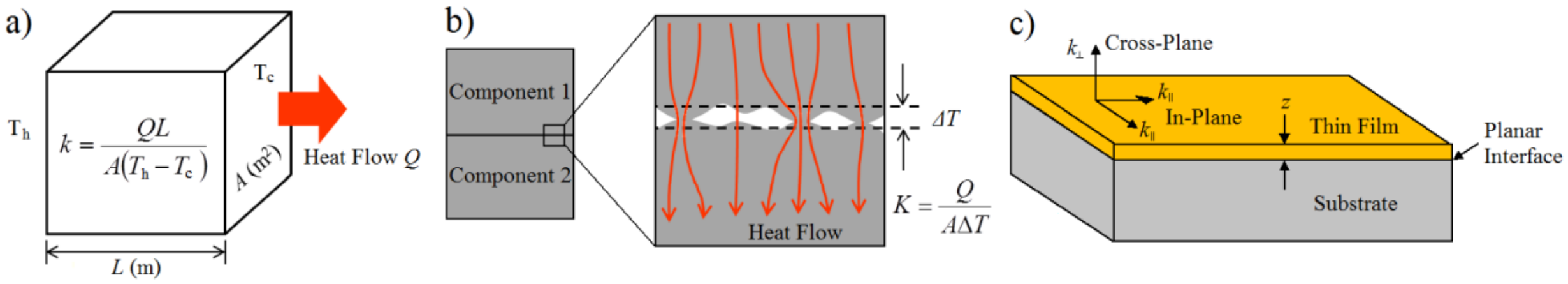

*Figure 1. a) Definition of thermal conductivity of a solid bulk material using Fourier's law of heat conduction. Temperature at the left and right sides are $T_h$ and $T_c$, respectively. Heat transfer cross-sectional area is denoted by $A$. The heat flow is assumed to be one-dimensional with no lateral heat loss. b) Definition of interfacial thermal conductance $K$ at the contact between two bulk materials. Bulk materials are contacted through limited contact points, which results in a temperature drop $\Delta T$ across the interface. c) Schematic of a typical thin film material on a substrate structure. The thermal conductivity of thin films is in general anisotropic due to the geometric constraints, which means that the value are different in different directions (i.e. in-plane thermal conductivity $k_\parallel$, and cross-plane thermal conductivity $k_\perp$). Due to the difference in materials, there usually exists a thermal boundary resistance (the inverse of interfacial thermal conductance) at the bonded planar interface.*

Extensive efforts have been made since the 1950s for the characterization of thermal conductivity and thermal contact resistance in bulk materials [3–8]. Table 1 summarizes some of the most commonly used measurement techniques, which in general can be divided into two categories: steady-state methods and transient methods. The steady-state methods measure thermal properties by establishing a temperature difference that does not change with time. Transient techniques usually measure time-dependent energy dissipation process of a sample. Each of these techniques has its own advantages and limitations, and is suitable for only a limited range of materials, depending on the thermal properties, sample configuration, and measurement temperature. Section 2 is devoted to comparing some of these measurement techniques when applied for bulk materials.

Thin film form of many solid materials with a thickness ranging from several nanometers to hundreds of microns have been extensively used in engineering systems to improve mechanical, optical, electrical and thermal functionality, including microelectronics [9], photonics [10], optical coating [11], solar cells and thermoelectrics [12]. Thin film materials can be bonded on a substrate (Figure 1c), free-standing, or in a multilayer stack. When the thickness of a thin film is smaller than the mean free path of its heat carriers, which are electrons and phonons depending on whether the material is electrically conducting or not, the thermal conductivity of thin films is reduced compared to its bulk counterparts because of the geometric constraints. Thermal conductivity of thin films is usually thickness-dependent and anisotropic, where the heat conducting capability in the direction perpendicular to the film plane (cross-plane) is very different from that parallel to the film plane (in-plane), as shown in Figure 1c. The thermal conductivity of thin films also depends strongly on the materials preparation (processing) method, and the substrate that thin films are sitting on. The conventional thermal conductivity measurement techniques for bulk materials are usually too large in size to measure the temperature drop and the heat flux across a length scale ranging from a few nanometers to tens of microns. For example, the smallest beads of commercial thermocouples have a diameter of around 25 μm, which could be much larger than the thicknesses of most electronic thin films.

*Table 1. Commonly used thermal characterization techniques reviewed in this article.*

| | **Bulk material** | **Thin film** |
|---|---|---|
| **Steady-state** | Absolute technique;<br>Comparative technique;<br>Radial heat flow method;<br>Parallel conductance method | Steady-state electrical heating methods |
| **Transient (frequency-domain)** | Pulsed power technique | 3ω method;<br>FDTR technique |
| **Transient (time-domain)** | Hot-wire method (needle-probe method);<br>Laser flash method;<br>Transient plane source (TPS) method | TDTR technique |

Significant progresses have also been made for the characterization of thermal conductivity and thermal boundary resistance of thin films over the past 30 years due to the vibrant research in micro- and nanoscale heat transfer [13–19]. Section 3 reviews a few measurement techniques for thin films including the steady-state methods, the $3\omega$ method, and the transient thermoreflectance technique in both time-domain (TDTR) and frequency-domain (FDTR), as summarized in Table 1. We note that some techniques (e.g. 3ω, TDTR and FDTR) are actually very versatile and can be applied for the thermal characterization of both bulk and thin film materials although the techniques reviewed here have been divided into two categories for bulk materials and thin films just for convenience.

## 2. Bulk materials

### *2.1 Steady-state methods*

In the steady-state measurement, the thermal conductivity and interfacial thermal conductance are determined by measuring the temperature difference $\Delta T$ at a separation (distance) under the steady-state heat flow $Q$ through the sample. Figure 2 shows the schematic of four different steady-state methods commonly adopted: absolute technique, comparative cut bar technique, radial heat flow method, and the parallel thermal conductance technique.

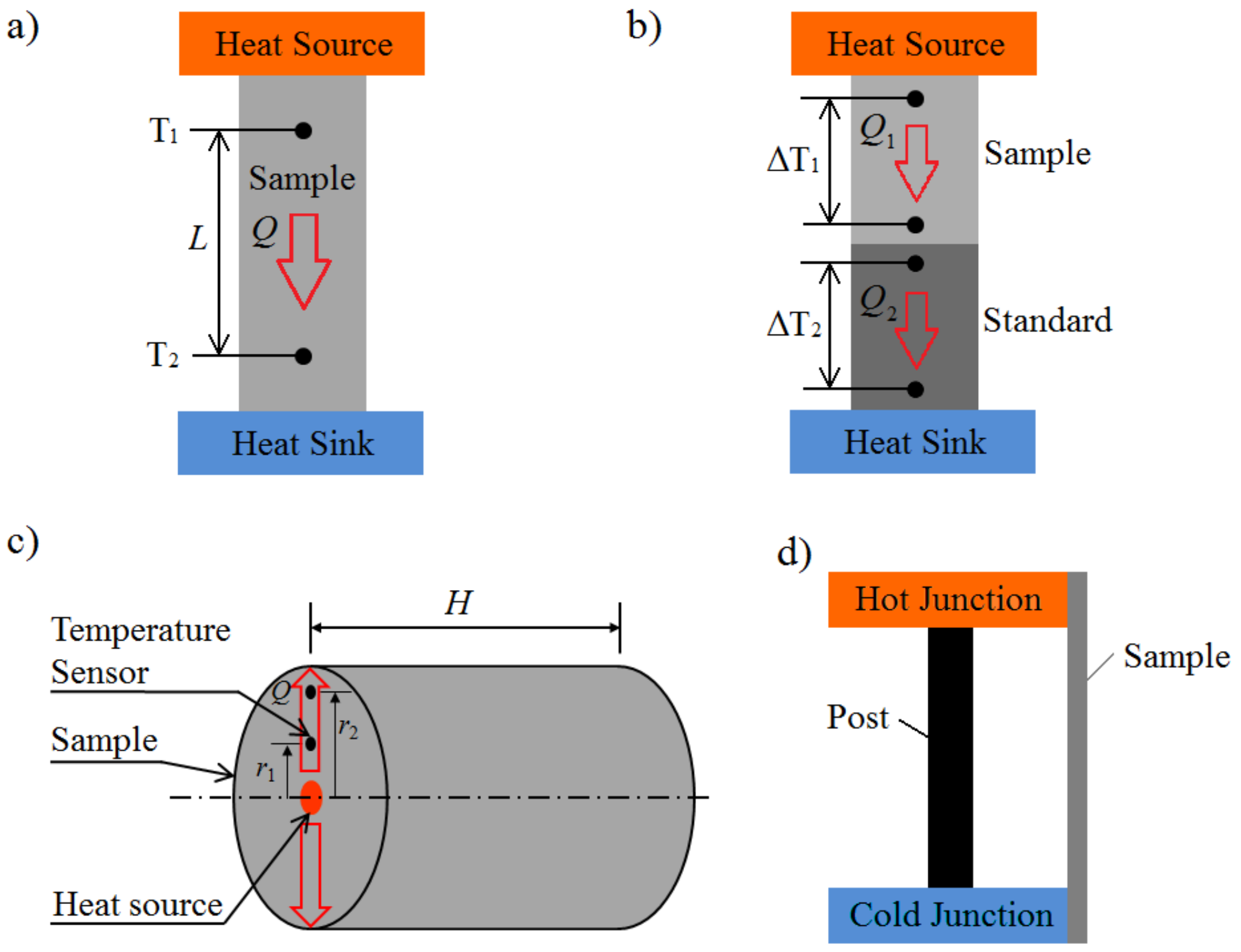

*Figure 2. Schematic of steady-state methods commonly used for measuring the thermal conductivity of bulk materials. a) Absolute technique. The sample is placed between a heat source and a heat sink, with known power output, which results in temperature drop across a given length (separation) of the sample. b) Comparative cut bar technique. A standard material with known thermal conductivity is mounted in series with the test sample. c) Radial heat flow method. A cylindrical sample heated from its axis and as heat flows radially outwards, a steady-state temperature gradient in the radial direction is established. d)*

*Parallel thermal conductance technique for measuring small needle-like samples that cannot support heaters and thermocouples. A sample holder or stage is used between the heat source and heat sink. A thermal conductance measurement for the sample holder is carried out first to quantify the heat loss associated with the sample holder. The testing sample is then attached to the sample holder and the thermal conductance is measured again. Thermal conductivity of the measured sample can then be deduced by taking the difference of these two measurements.*

### *2.1.1 Absolute technique*

Absolute technique is usually used for samples that has a rectangular or cylindrical shape. When conducting this measurement, the testing block is placed between a heat source and a heat sink as shown in Figure 2a. The sample is heated by the heat source with known steady-state power input and the resulting temperature drop $\Delta T$ across a given length (separation) of the sample is measured by temperature sensors after a steady-state temperature distribution is established. The temperature sensors employed can be thermocouples and thermistors. Thermocouples are the most widely used sensors due to their wide range of applicability and accuracy. The resulting measurement error in $\Delta T$ due to temperature sensor shall be less than 1% [20]. Thermal conductivity $k$ of the sample can be calculated using Fourier's law of heat conduction:

$$k = \frac{QL}{A\Delta T} \quad (1)$$

$$Q = p - Q_{\text{loss}} \quad (2)$$

where $Q$ is the amount of heat flowing through the sample, $A$ is the cross-sectional area of the sample, $L$ and $\Delta T$ are the distance and temperature difference between temperature sensors, $p$ is the applied heating power at heat source side, and $Q_{\text{loss}}$ is the parasitic heat losses due to radiation, conduction and convection to the ambient.

The major challenge of the absolute technique is to determine the heat flow rate $Q$ through the sample at the presence of parasitic heat losses $Q_{\text{loss}}$ and to measure temperature difference $\Delta T$ accurately. Parasitic heat losses include convection and radiation to the surrounding and conduction through thermocouple wires. In general, parasitic heat losses should be controlled to be less than 2% of the total heat flow through the sample. To minimize convection and radiation heat losses, most measurements are conducted under vacuum with radiation shields [21]. Besides the convection and radiation heat losses, another concern is that the heat conduction through thermocouple wires. It is therefore preferable to use thermocouples with small wire diameter (e.g. 0.001 inch [3]) and low thermal conductivity wires (e.g. chromel-constantan). Also, in order to minimize the conduction heat loss, differential thermocouple with only two wires coming off the sample can be applied to directly obtain the temperature difference $\Delta T$ [3]. A typical test apparatus of the absolute technique is the guarded-hot-plate apparatus. ASTM C177 [20], European Standard EN 12667 [22] and International Standard ISO 8302 [23] have more details about the apparatus and testing procedure. Major drawbacks of the absolute technique include: 1) the testing sample should be relatively large (in centimeter scale or even larger) and can form a standard circular or rectangular shapes when thermocouples are used for measuring temperatures. 2) the test usually suffers from a long waiting time, up to a few hours, to reach steady state. Resistance thermometers (i.e. RTDs) [24,25] and infrared (IR) thermography [26] are often employed for temperature sensing when testing small samples (in micron scale or even smaller) by using absolute technique.

The absolute technique has also been applied for measuring the thermal contact resistance between two components, as shown in Figure 3a. Testing samples are pressed together with controllable contact pressures. Several thermocouples (usually 4 or 6) are placed inside the two mating samples at uniform intervals (separations) to measure local temperature in response to an applied heating power. Once the steady-state heat flow is achieved, temperatures are recorded and plotted (denoted by solid dots) in the temperature versus distance curves as depicted in Figure 3b. Temperatures at both interfaces (i.e. $T_h$ and $T_c$, denoted by hollow circles) can be deduced assuming the temperature distribution on each side is linear. The thermal contact resistance is then calculated by the temperature drop (i.e. $T_h$ - $T_c$) divided by total heat flow across the interface. In order to obtain an accurate thermal contact resistance, the temperature drop across the interface should be maintained relative large (e.g. > 2 °C [27]) through the control of applied heating power.

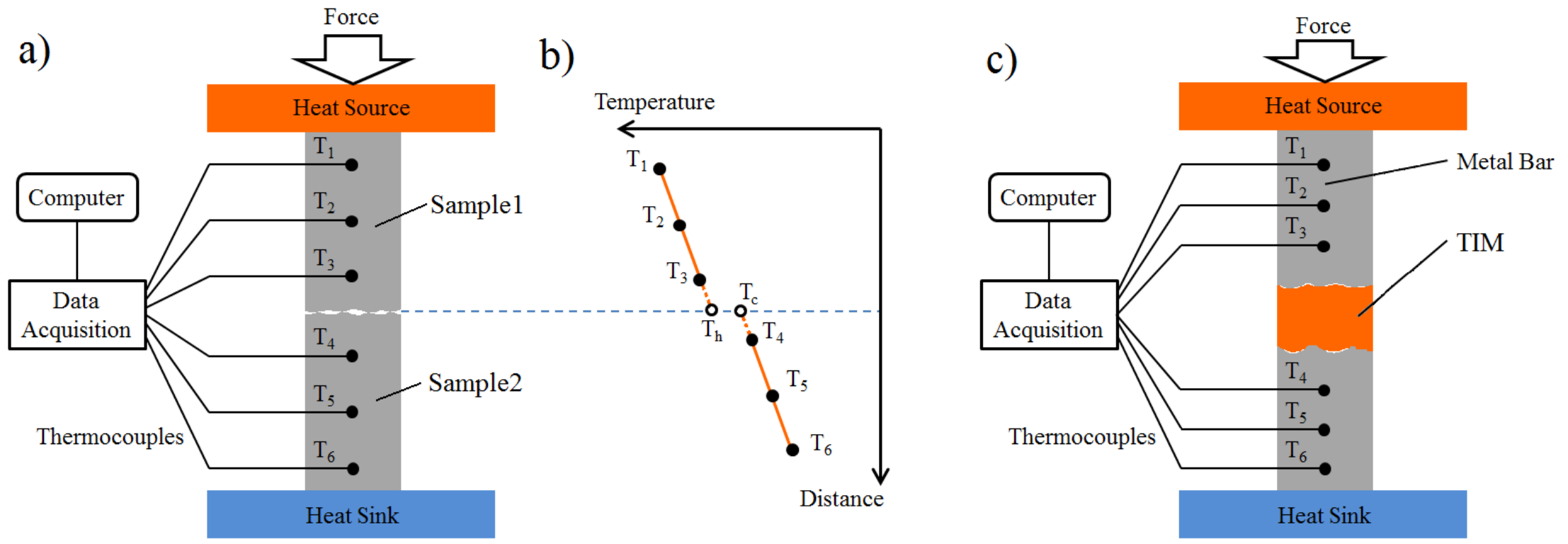

*Figure 3. a) Measurement configuration and b) an illustration of data processing for thermal contact resistance of bulk materials. Thermal contact resistance is calculated by dividing the temperature drop ($T_h$ - $T_c$) across the interface by the heat flow. c) Measurement configuration for thermal interface materials (TIMs) according to ASTM D5470 standard (not to scale).*

In addition to bare contacts, thermal interface materials (TIMs) are usually employed to reduce thermal contact resistance. Commonly used TIMs include pressurized gases (e.g. helium, hydrogen), thermal greases (e.g. silicone oil, glycerin), thermal adhesive, thermal conductive pad and polymer matrix composited with high thermal conductive particulates such as silver and ceramics as fillers, phase change material, and solders [28]. The one-dimensional steady-state testing method is widely used for characterizing the thermal conductivity of TIMs, as well as the thermal contact resistance [29]. The testing instrument is similar to Figure 3a, except that the two testing samples are replaced by two metal blocks (usually copper) with known thermal conductivity and the TIM to be investigated is inserted in between as shown in Figure 3c.

*2.1.2 Comparative technique*

The biggest challenge in the absolute technique is to accurately determine the heat flow through sample. However, if one has a standard material whose thermal conductivity is known, the comparative cut bar technique can be applied and the direct measurement of heat flow is unnecessary. Figure 2b shows the measurement configuration of the comparative cut bar technique, which is similar to that of the absolute method. At least two temperature sensors should be employed on each bar. Extra sensors can be used to conforming linearity of temperature versus distance along the column. The selection of temperature sensors depends on the system size, temperature range, meter bars, specimen and gas within the system [30], while thermocouples are the most widely employed temperature sensors. Since the amount of heat flow through the standard material equals to the heat flow through the measurement sample target, the thermal conductivity of the measurement sample target is given by:

$$k_1 = k_2 \frac{A_2 \Delta T_2 L_1}{A_1 \Delta T_1 L_2} \quad (3)$$

where subscript 1 and 2 are associated with the sample and the standard material, respectively.

By implementing the standard material with known thermal conductivity, the sample target thermal conductivity measured can be extracted without heat flow measurement as shown in Eq. (3), and the associated error due to heat flow measurement can thus be eliminated. However, efforts are still needed to ensure equal heat flow between the standard material and the testing specimen. This technique achieves the best accuracy when the thermal conductivity of the measurement target is comparable to that of the standard material [3]. This is the most widely used method for axial heat flow thermal conductivity testing. ASTM E1225 [30] gives the experimental requirements and procedure for the comparative cut bar technique.

Another type of comparative technique is the heat flow meter method. A heat flux transducer is used, which essentially replaces the standard material in the comparative cut bar method. By appropriate calibration of the heat flux transducer using a specimen of known thermal conductivity, the thermal conductivity of measurement sample can then be easily determined using Fourier's law of heat conduction with measured heat flux. This method is usually used to characterize low thermal conductivity materials, such as building insulation materials. ASTM C518 [31], ASTM E1530 [32] and European Standard EN 12667 [22] specify the test apparatus and calibration method for heat flow meter method.

### *2.1.3 Radial heat flow method*

The two steady-state methods described above use a longitudinal arrangement of samples to measure its thermal conductivity. This can be satisfactory at low temperatures. However, for measurement at very high temperatures (e.g. > 1000 K), radiation heat loss from the heater and sample surfaces is not negligible and can cause large uncertainties when quantifying the heat flow through the sample. In order to overcome this, samples with cylindrical geometry are used in the radial heat flow method. Flynn described the evolution of the apparatus used for measurement of thermal conductivity by using this technique [33]. The ASTM C335 [34] and the ISO 8497 [35] cover the relevant measurement requirements and testing procedure using this method. The sample is heated internally at the axis and the heat flows radially outwards as depicted in Figure 2c. A steady-state temperature gradient in the radial direction is established. Thermocouples are used predominantly for temperature sensing in radial heat flow method, with an accuracy within ±0.1 °C [34]. The thermal conductivity is derived from Fourier's law of heat conduction in cylindrical coordinate:

$$k = \frac{Q\ln(r_2/r_1)}{2\pi H \Delta T} \tag{4}$$

where $r_1$ and $r_2$ are the radius where the two temperature sensors are positioned, $H$ is the sample height and $\Delta T$ is the temperature difference between the temperature sensors.

### *2.1.4 Parallel thermal conductance technique*

Characterization of small bulk materials with a size in the millimeter scale is very challenging because temperature sensing by thermocouples and the heat flux measurement are extremely difficult. The parallel thermal conductance technique was introduced by Tritt *et al* [36] for small needle-like samples (e.g. 2.0 × 0.05 × 0.1 $mm^3$ [36], 10 × 1 × 1 $mm^3$ [37]). Figure 2d shows the typical experimental configuration, which is a variation of the absolute technique for those samples that cannot support heaters and thermocouples. A sample holder or stage is used between the heat source and heat sink. Differential thermocouple is positioned between the hot side and the post on the one junction end, and between the cold side and the post on the other junction. Before measuring the thermal conductivity of the specimen, a thermal conductance measurement of the sample holder is performed first to quantify the thermal losses associated with the sample holder. The testing sample is then attached to the sample holder and the thermal conductance is measured again. Thermal conductance of the sample can be deduced by taking the difference of these two measurements. Thermal conductivity is then calculated from the thermal conductance by multiplying sample length and dividing by the sample cross-sectional area. The major drawback of this method is the requirement to measure cross-sectional area of such small samples. Inaccuracies in cross-sectional area measurement can lead to large uncertainties in the calculated thermal conductivity.

*2.2 Transient technique*

In order to overcome the drawbacks associated with the steady-state methods described above such as parasitic heat losses, contact resistance of temperature sensors, and long waiting time for establishing steady-state temperature difference, a variety of transient techniques have been developed. The heat sources used in transient techniques are supplied either periodically or as a pulse, resulting in periodic (phase signal output) or transient (amplitude signal output) temperature changes in the sample, respectively. This section focuses on the four commonly used transient techniques, namely pulsed power technique, hot-wire method, transient plane source (TPS) method and laser flash thermal diffusivity method.

*2.2.1 Pulsed power technique*

Pulsed power technique was first introduced by Maldonado to measure both thermal conductivity and thermoelectric power [38]. This technique is a derivative of the absolute technique in the steady-state methods, with the difference that a periodic electrical heating current $I(t)$ is used. This technique is in principle very close to the Angstrom's method [39,40] in terms of the heating method. But the difference is that the heat sink temperature of this technique is slowly varying during the measurement. Figure 4a shows the schematic of a typical setup for pulsed power technique. The sample (usually in cylindrical or rectangular geometry) is held between a heat source and a heat sink. The heating current used can be either a square wave of constant-amplitude current or a sinusoid wave [41]. During the experiment, a periodic electric current with a period of $2\tau$ is applied to the heat source while the temperature of the heat sink bath $T_c$ drift slowly. A small temperature gradient $\Delta T = T_h - T_c$ (usually ~0.3 K) is created between the heat source and the heat sink, which can be measured by a calibrated Au-Fe chromel thermocouple [38]. The heat balance equation between the heat dissipated by the heater and conducted through the sample is given as:

$$Q = C(T_h)\frac{dT_h}{dt} = R_e(T_h)I^2(t) - K\left(\frac{T_c + T_h}{2}\right)\Delta T(t) \qquad (5)$$

where $R_e(T_h)$ is the electrical resistance of the heater which changes with temperature.

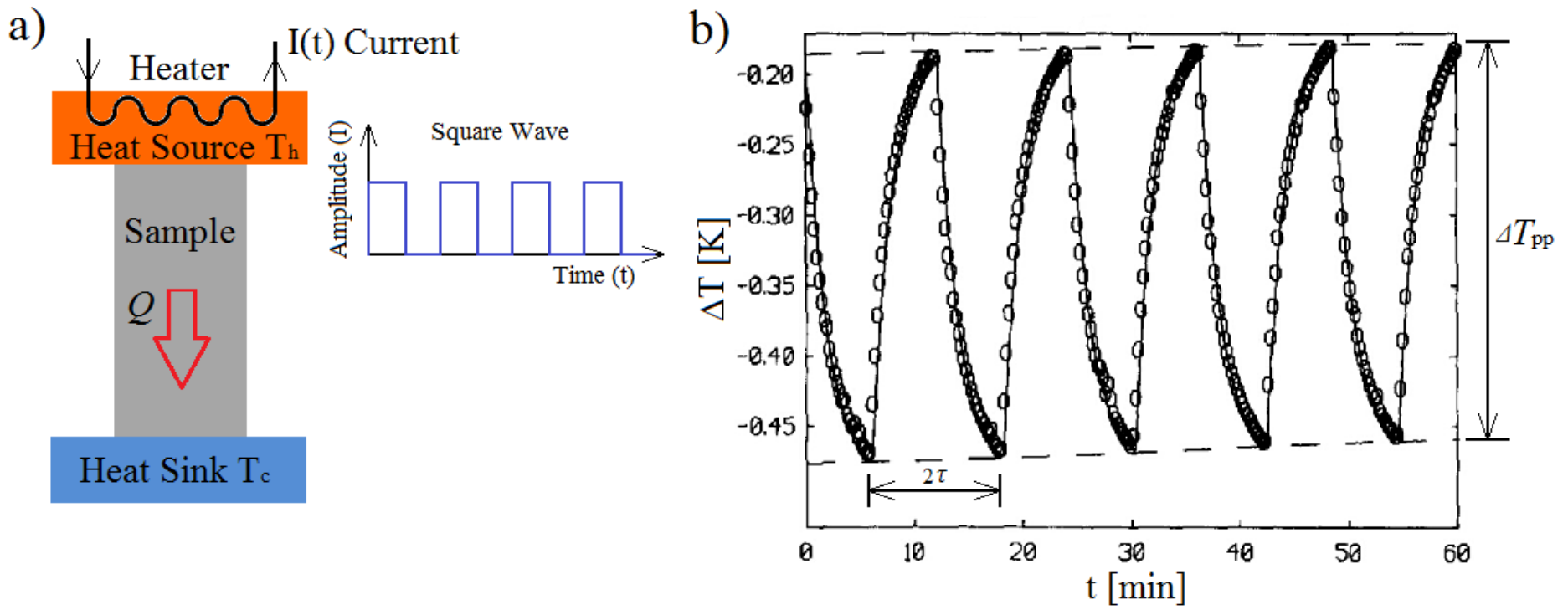

*Figure 4. a) Schematic diagram of a typical pulsed power technique for thermal conductivity measurement where periodic square-wave current is applied. b) The time-dependent temperature difference $\Delta T = T_h - T_c$ across the sample, from Ref. [38] where the temperature of the heat sink is slowly drifted. The dots represent experimental data and line represents calculation results.*

It is possible to obtain thermal conductance $K$ as a function of temperature from the measured temperature $T_h$. However, Eq. (5) is nonlinear and difficult to be solved analytically. Therefore, $C(T_c)$, $R(T_c)$ and $K(T_c)$ are usually used in Eq. (5) to replace $C(T_h)$, $R(T_h)$ and $K\left((T_h + T_c)/2\right)$ to linearize the equation. This assumption holds because temperature difference between $T_c$ and $T_h$ (i.e. $\Delta T$) is very small (Figure 4b). Also, $T_c$ is considered to be constant since it is drifted very slowly compared to the periodic current. The final solution has an oscillating sawtooth-like shape as shown in Figure 4b. Smooth curves (i.e. the two dashed lines) are drawn through the maxima and minima of the oscillations. The difference between the two dashed smooth curves $\Delta T_{pp}$ yields a relation for the thermal conductance of the measured sample [38]:

$$K = \frac{RI_0^2}{\Delta T_{pp}} \tanh\left(\frac{K\tau}{2C}\right) \tag{6}$$

where $\tau$ is the half period of the heating current, $C$ is the volumetric heat capacity, $R$ is thermal resistance and $I_0$ is the amplitude of electric current.

Numerical iteration can be applied to solve for the thermal conductance term $K$ in Eq. (6) since all other parameters are known as a function of temperature. This technique is capable of measuring a wide temperature range from 1.9 to 390 K as reported in literature [42–44], and an ultralow thermal conductivity as low as 0.004 $\mathrm{W/m{\cdot}K}$ for $ZrW_2O_8$ at temperature 2 K [43]. The measurement uncertainty reported by Maldonado is less than 3% [38].

*2.2.2 Hot-wire method*

The hot-wire method is a transient technique that measures temperature rise at a known distance from a linear heat source (i.e. hot wire, usually platinum or tantalum) embedded in the test sample. Stalhane and Pyk employed this method in 1931 to measure the thermal conductivity of solids and powders [45]. The method assumes an idealized “one-dimensional radial heat flow” inside the isotropic and homogeneous test sample, which is based on the assumption that the linear heat source has infinite length and infinitesimal diameter as shown in Figure 5. When there is an electric current of constant intensity passes through the hot wire, thermal conductivity of the test sample can be derived from the resulting temperature change at a known distance from the hot wire over a known time interval. Hot-wire method is commonly used to measure low thermal conductivity materials such as soils [46], ice cores [47] and refractories (refractory brick, refractory fibers, plastic refractories and powdered materials [48]). This method has also been commonly used for measuring the thermal conductivity of liquids. ASTM C1113 and ISO 8894 specify more details on apparatus and test procedure for the measuring of refractories using hot-wire method [48,49].

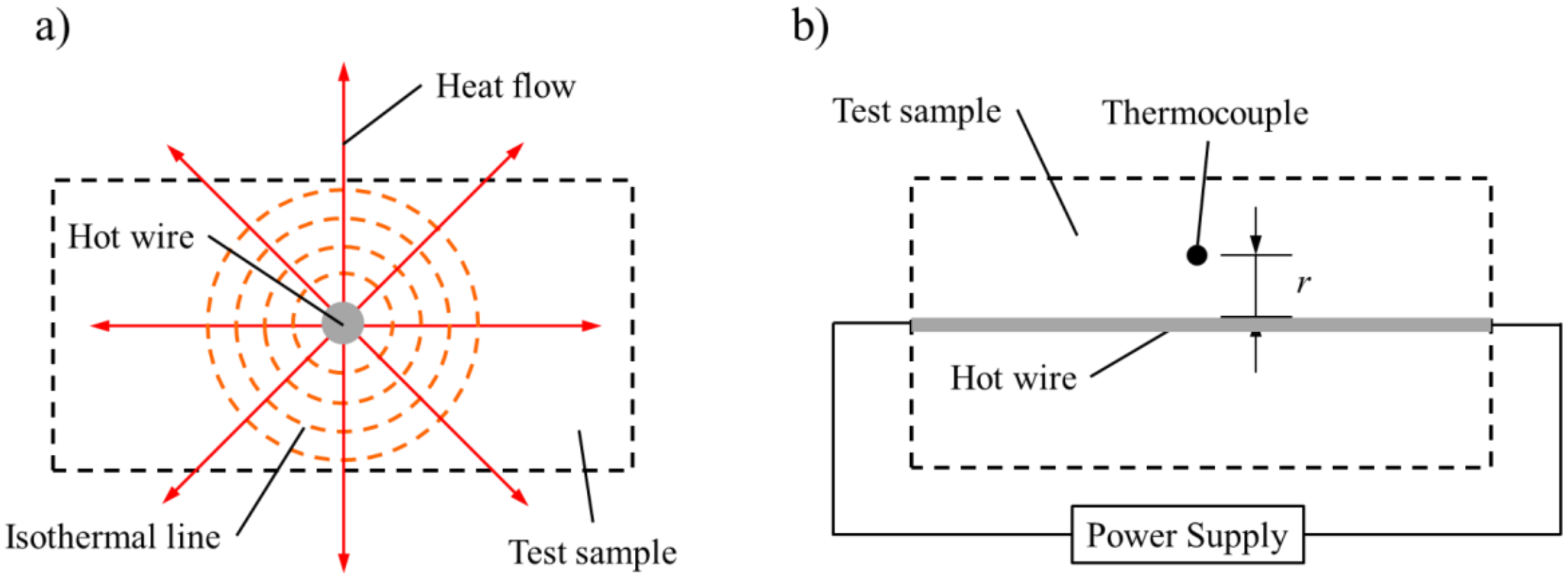

*Figure 5. The measurement principle of the hot-wire method. a) One-dimensional radial heat flow is assumed inside the test sample. b) As an electric current of fixed intensity passes through the hot wire, the thermal conductivity of the test sample can be derived from the resulting temperature change at a known distance from the hot wire over a known time interval. This figure is modified from Ref. [50].*

As the hot wire produces a thermal pulse for a finite time with constant heating power, it generates isothermal lines in an infinite homogeneous medium initially at thermal equilibrium. The transient temperature for sufficiently long time from the start of the heating, can be expressed with good approximation by [50]:

$$T(r,t) = \frac{p}{4\pi k L}\left[\ln\left(\frac{4\alpha t}{r^2}\right) + \frac{r^2}{4\alpha t} - \frac{1}{4}\left(\frac{r^2}{4\alpha t}\right) - \cdots - \gamma\right] \tag{7}$$

For sufficiently large values of the time $t$, the terms $r^2/4\alpha t$ inside the parenthesis is negligible because it is far less than 1. The above equation can then be simplified to:

$$T(r,t) = \frac{p}{4\pi kL}\left[\ln\left(\frac{4\alpha t}{r^2}\right) - \gamma\right] \tag{8}$$

The temperature rise at a point in the test sample from time $t_1$ to $t_2$ is given by:

$$\Delta T = T(t_2) - T(t_1) = \frac{p}{4\pi kL}\ln\left(\frac{t_2}{t_1}\right) \tag{9}$$

Thermal conductivity is then obtained from the temperature rise $\Delta T$ versus natural logarithm of the time $\ln(t)$ expressed below:

$$k = \frac{p}{4\pi[T(t_2) - T(t_1)]L}\ln\left(\frac{t_2}{t_1}\right) \tag{10}$$

It should be noted that when $r$ in Eq. (8) equals to 0, the wire will act both as a line source heater and a resistance thermometer. Today's hot-wire method instruments allow more than 1000 data readings of the transient temperature rise from times less than 1ms up to 1s (or 10s, in the case of solids) coupled with finite element methods to establish a very low uncertainty [51]. If applied properly, it can achieve uncertainties below 1% for gases, liquids, and solids, and below 2% for nanofluids [51]. Despite its advantages, there are very few commercial hot-wire instruments [52]. Possible reason is due to the delicacy of the very thin wire which easily snaps.

The needle-probe method, also for testing isotropic and homogeneous materials, is a variation of the hot-wire method. Working principle of the needle-probe method is same as the hot-wire method. But the temperature sensing is based on a zero radius system (i.e. $r$ = 0 in Eq. (8)). The heating wire and the temperature sensor (thermocouple) are encapsulated in a probe that electrically insulates the heating wire and the temperature sensor from the test sample. The probe helps protect the heating wire. This configuration is particularly practical where thermal conductivity is determined by a probe inserted in the test sample. Therefore, the method is conveniently applied to powder-like materials and soils. A needle-probe device can be used to measure sample thermal properties in situ, but most commonly a temperature-controlled furnace is used to produce the base temperatures for the measurements [53]. ASTM D5930 [54] and ASTM D5334 [55] standardize the test procedure and data analysis method for the needle-probe method.

*2.2.3 Transient plane source (TPS) method*

Transient plane source (TPS) method (i.e. hot disk method) uses a thin metal strip or a disk as both a continuous plane heat source and a temperature sensor as depicted in Figure 6a. The metal disk is first sealed by electrical insulation and then sandwiched between two identical thin slab-shaped testing samples. All other surfaces of the testing samples are thermally insulated. During the experiment, a small constant current is applied to the metal disk to heat it up. Since the temperature increase of the metal disk is highly dependent on the two testing samples attached to it, thermal properties of the testing samples can be determined by monitoring the temperature increase for a short time period. This time period is generally only a few seconds so that the metal disk can be considered in contact with infinite size samples throughout the transient signal recording process. Temperature increase at the sensor surface $\Delta T$ (e.g. 1-3°C [56]) as a function of time can be measured. Measurement accuracy of the temperature sensor (temperature resistance thermometer) is usually ± 0.01 °C [56]. Then, fitting the temperature curves by Eqs. (11) and (12) to the measured $\Delta T$ renders the inverse of thermal conductivity $1/k$ [57].

$$\Delta T(\phi) = \frac{Q}{\pi^{1.5} r k} D(\phi) \tag{11}$$

$$\phi = \sqrt{\frac{t\alpha}{r^2}} \tag{12}$$

where $r$ is the sensor radius, $D(\phi)$ is a dimensionless theoretical expression of the time dependent increase describes heat conduction of the sensor (Figure 6b).

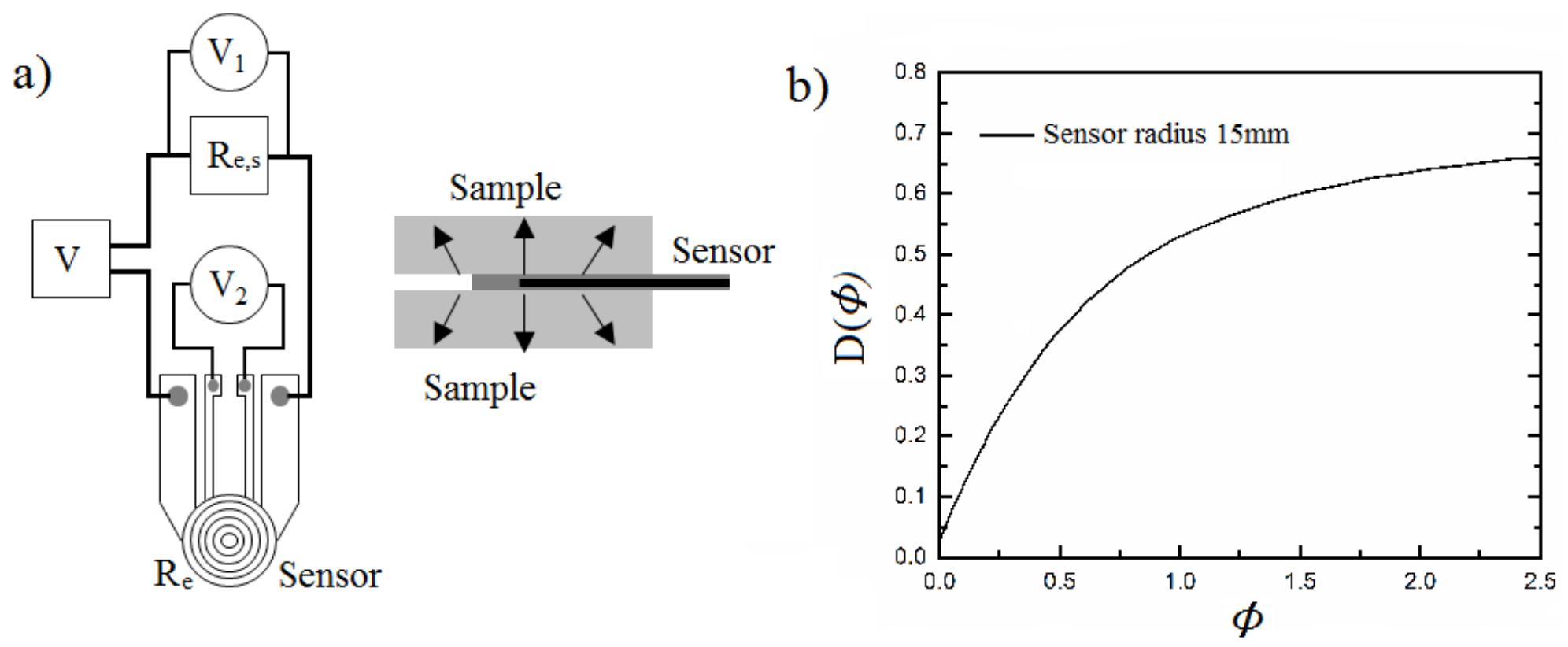

*Figure 6. a) Top and cross-sectional views of the TPS method [58]. $V$ is a constant voltage source, $V_1$ and $V_2$ are precision voltmeters, $R_{e,s}$ is a standard resistor and $R_e$ is the nonlinear resistance of the sensor. The metal disk (i.e. sensor) is sealed by electrical insulation and then sandwiched between two test samples. A small constant current is applied to the metal disk to heat it up. Since the temperature increase of the metal disk is highly dependent on the two testing samples attached to it, thermal properties of the testing samples can be determined by monitoring this temperature increase for a short time period. b) A typical plot shows $D(\phi)$ as a function of $\phi$. $\phi$ is defined in Eq. (12). $D(\phi)$ is a dimensionless theoretical expression of the time dependent increase that describes heat conduction of the sensor. The curve is reconstructed from data published in Ref [57].*

The TPS method was reported to be capable for measuring materials with thermal conductivity ranging from 0.005 to 500 $\mathrm{W/m \cdot K}$ in the temperature range from cryogenic temperatures to 500 K, including liquids, aerogels and solids [59–62]. ASTM D7984 [56] and ISO 22007-2 [63] specify the test apparatus and procedure for the TPS method. One drawback of TPS measurement is that each of the two sample pieces needs to have one entirely planar side. This makes it difficult for some materials, especially powders or granules [64]. The TPS measurement errors come from several sources: 1) thermal contact resistance between the sensor and the testing samples, 2) the thermal inertia of the sensor, 3) the measured power input being influenced by the heat capacity of the electrical isolation films, and 4) the electrical resistance change of the metallic disk sensors. Model corrections for these errors are necessary to improve measurement accuracy when doing data analysis. Readers can refer to Refs. [59,60,62] for more information. For example, when considering the influence of the sensor's electrical resistance change with temperature, Eq. (11) needs to be revised to [62]:

$$\Delta T(\tau) = \left[\frac{Q^2 V_1 V_2}{(V_1 + V_2)^2 R_{\mathrm{e,\,s}}} - \frac{C\Delta T(t)}{t}\right]\frac{D(\phi)}{\pi^{1.5} r k} \tag{13}$$

### *2.2.4 Laser flash method for thermal diffusivity*

Contact thermal resistance is a major source of error for temperature measurement. The laser flash method employs non-contact, non-destructive temperature sensing to achieve high accuracy [65]. The method was first introduced by Parker *et al*. [66]. It uses optical heating as an instantaneous heating source, along with a thermographic technique for quick non-invasive temperature sensing. The testing sample is usually a solid planar-shaped material when measuring thermal conductivity, and is a multi-layer structure when characterizing thermal contact resistance. A typical measurement configuration for laser flash method is depicted in Figure 7a. An instantaneous light source is used to uniformly heat up the sample's front side and a detector measures the time-dependent temperature rise at the rear side. Heat conduction is assumed to be one-dimensional (i.e. no lateral heat loss). The testing sample is usually prepared by spraying a layer of graphite on both sides to act as an absorber on the front side and as an emitter on the rear side for temperature sensing [67]. The rear-side infrared radiation thermometer should be fast enough to respond to the emitting signals, and the precision of temperature calibration is usually ± 0.2 K [53]. Dynamic rear-side temperature response curve (Figure 7b) is used to fit for the thermal diffusivity. The higher the thermal diffusivity of the sample, the faster heat transfer and temperature rise on the rear side.

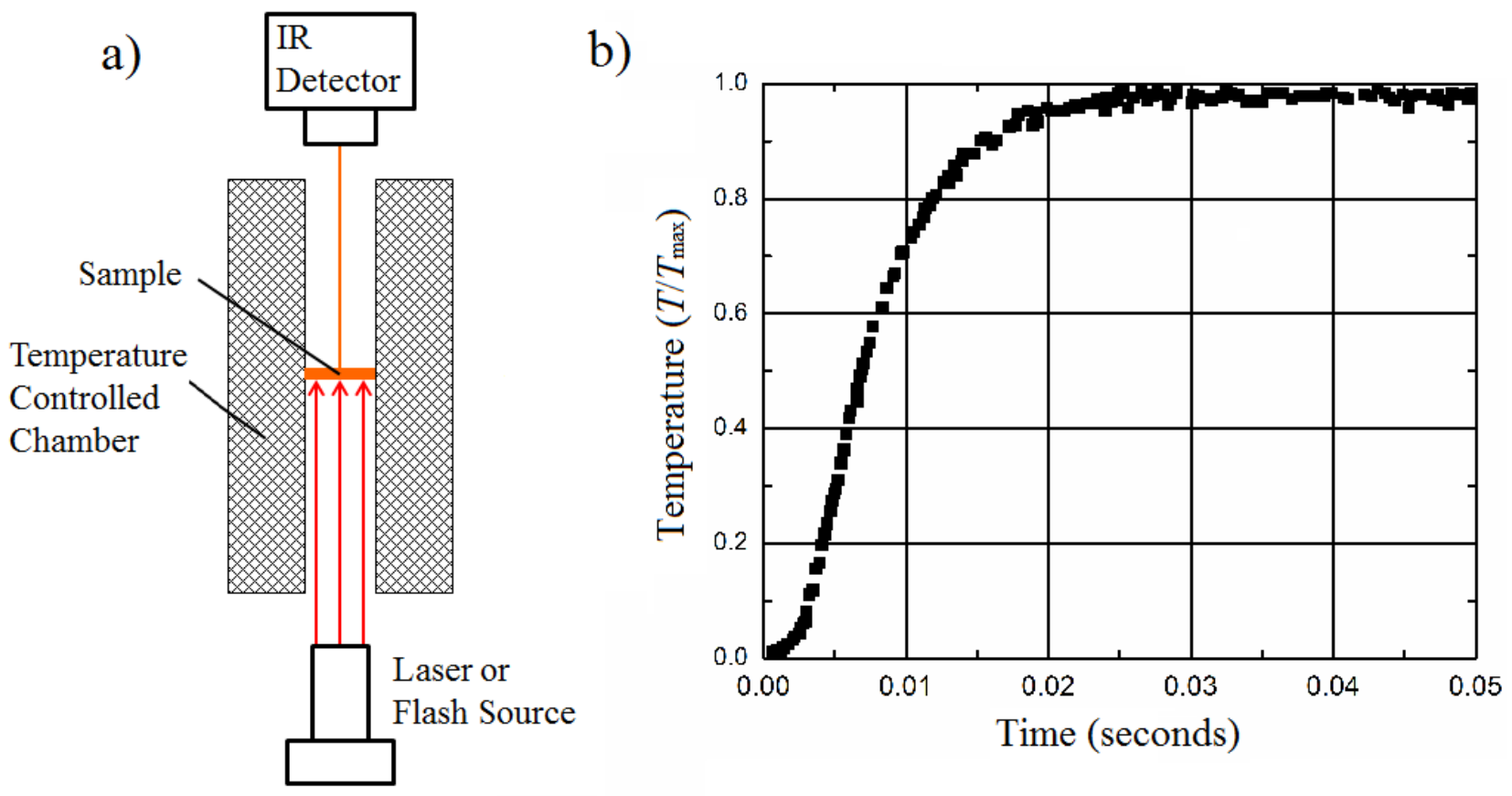

*Figure 7. a) Schematic of the laser flash method for thermal diffusivity. A light source heats the front side of the planar-shaped sample and an infrared detector measures the time-dependent temperature rise at the rear side. Thermal diffusivity of the sample is obtained by fitting the dynamic response of rear side temperature rise of the sample. b) An example showing the measured rear side temperature rising curve. The curve is reconstructed from data published in Ref* [68].

Theoretically, the temperature rise at the rear-side as a function of time can be written as [66]:

$$T(t) = \frac{\boldsymbol{q}}{\rho c_p d}\left[1 + 2\sum_{n=1}^{\infty}(-1)^n \exp\left(\frac{-n^2\pi^2}{d^2}\alpha t\right)\right] \qquad (14)$$

where $d$ is the sample thickness, and $\alpha$ is thermal diffusivity. To simplify Eq. (14), two dimensionless parameters, $W$ and $\eta$ can be defined:

$$W(t) = T(t)/T_{max} \qquad (15)$$

$$\eta = \pi^2\alpha t/d^2 \qquad (16)$$

$T_{max}$ denotes the maximum temperature at the rear-side. The combination of Eqs. (14-16) yields:

$$W = 1 + 2\sum_{n=1}^{\infty}(-1)^n \exp(-n^2\eta) \qquad (17)$$

When $W$ is equal to 0.5, which means the rear-side temperature reaches to one half of its maximum temperature, $\eta$ is equal to 1.38, and so the thermal diffusivity $\alpha$ is calculated by [66]:

$$\alpha = \frac{1.38d^2}{\pi^2 t_{1/2}} \qquad (18)$$

where $t_{1/2}$ is the time that takes for the sample to heat to one half maximum temperature on the rear surface.

ASTM E1461 [69] and ISO 22007-4 [70] specify the requirements of apparatus, test sample and procedure for thermal diffusivity measurement by the laser flash method. In addition to the thermal diffusivity $\alpha$ measured using the laser flash method, material density $\rho$ and specific heat $c_p$ need to be measured from separate experiments, to obtain the thermal conductivity using the relationship $k = \alpha\rho c_p$. The laser flash method is capable of measuring thermal conductivity over a wide temperature range (-120 °C to 2800 °C [71]) with measurement uncertainty reported to be less than 3% [72]. The advantages of the method are not only its speed (usually 1-2 s for most solid), but also its capability to use very small samples, *e.g.* 5-12 mm in diameter [53]. There are, however, some considerations that should be kept in mind before carrying out laser flash measurement. First of all, sample heat capacity and density should be known or determined from separate experiments, which may result in the “stack up” of uncertainties and lead to larger errors. Another criticism of the laser flash method is that the effect of sample holder heating could lead to significant error in the measurements if not accounted for properly [73]. Though laser flash method can be used to measure thin films, thickness of the measured sample is limited by the timescales associated with the heating pulse and the infrared detector. Typical commercial laser flash instruments can measure samples with a thickness of ~100 μm and above depending on the thermal diffusivity of the sample. For thin film sample with a thickness less than 100 μm, one needs to resort to the 3ω method or transient thermoreflectance techniques developed over the past two decades.

## 3. Thin films

### 3.1 Steady-state methods

#### 3.1.1 Cross-plane thermal conductivity measurement

Temperature drop across a thin film sample needs to be created and measured to characterize the cross-plane thermal conductivity. Creating and measuring the temperature drop is extremely challenging when the sample thickness is as small as a few nanometers to tens of microns. Figure 8 shows the schematic of two steady-state measurement configurations being frequently employed. In both configurations, thin films with thickness $d_f$ are grown or deposited onto a substrate with high thermal conductivity and small surface roughness (e.g. polished silicon wafer). A metallic strip with length $L$ and width $2a$ ($L \gg 2a$) is then deposited onto the thin film whose thermal conductivity is to be determined. The metallic strip should have a high temperature coefficient of resistance, such as Cr/Au film. During the experiment, the metallic strip is heated by a direct current (DC) passing through it. The metallic strip serves as both an electrical heater and a sensor to measure its own temperature $T_h$.

The temperature at the top of the film $T_{f,1}$ is generally assumed to be the same as the average heater temperature $T_h$. The most straightforward way would be using another sensor to directly measure the temperature $T_{f,2}$ at the bottom side of the film (Figure 8a). But this approach complicates the sample preparation processes which usually involve cleanroom microfabrication. The other approach is using another sensor situated at a known distance away from the heater/sensor to measure temperature rise of the substrate right underneath it (Figure 8b). A two-dimensional heat conduction model is then used to infer the substrate temperature rise at the heater/sensor location from the measured substrate temperature rise at the sensor location [15].

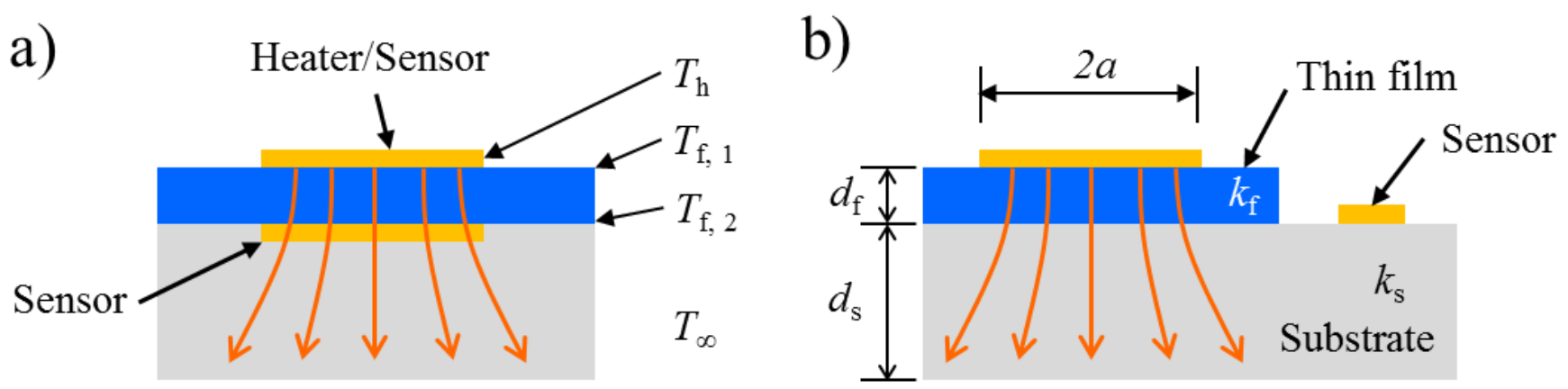

*Figure 8. Steady-state methods for measuring the cross-plane thermal conductivity of thin films. In both measurements, the metallic strip serves as both an electrical heater and a temperature sensor measuring its own temperature rise. The temperature $T_{f,1}$ on the top side of the thin film is generally assumed to be the same as the heater/sensor average temperature $T_h$. $T_\infty$ is the ambient temperature. In order to determine the temperature $T_{f,2}$ at the bottom of the thin film, either a) a sensor is deposited between thin film and substrate for direct measurement, or b) a sensor is deposited at a known distance away from the heater. A two-dimensional heat conduction model is then used to calculate the temperature at the bottom of the thin film.*

### 3.1.2 In-plane thermal conductivity measurement

The major challenge of measuring in-plane thermal conductivity is the evaluation of heat flow along the film in the presence of parasitic heat loss through the substrate. In order to increase measurement accuracy, Volklein *et al*. [74] concluded that it is desirable to have the product of the thin film in-plane thermal conductivity $k_{f,\parallel}$ and film thickness $d_f$ equal to or greater than the corresponding product of the substrate (i.e. $k_{f,\parallel} d_f \geq k_S d_S$). However, in order to completely remove parasitic heat loss through the substrate, a suspended structure by removing the substrate as shown in Figure 9 is desirable, which complicates microfabrication for sample preparation [75].

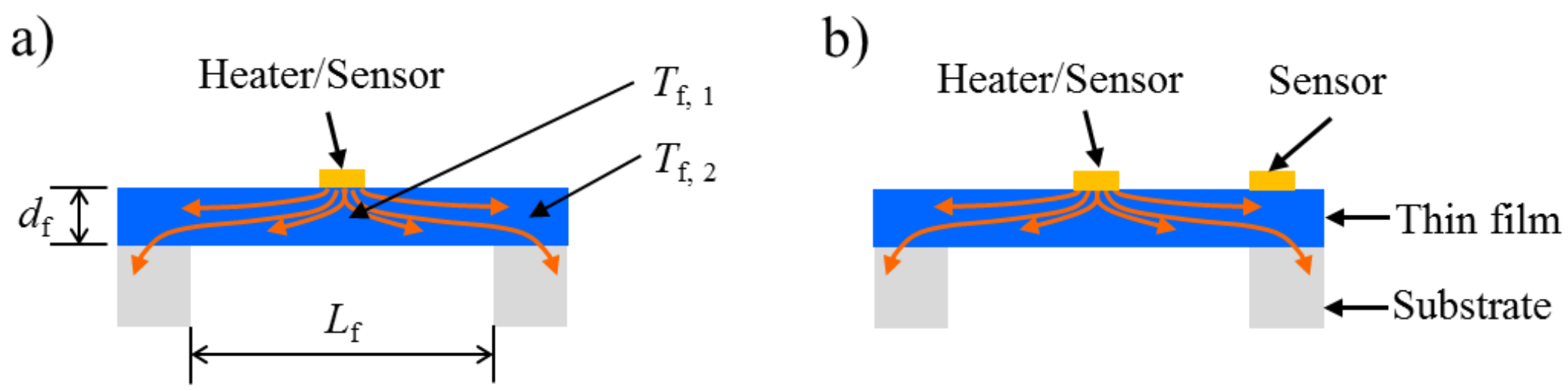

*Figure 9. a) In-plane thermal conductivity can be deduced from the difference in heater/sensor temperature rise of two measurements using two different thin film widths $L_f$ with all other parameters unchanged [74]. b) Steady-state thermal conductivity measurement with a heater/sensor and an additional temperature sensor.*

Figure 9 depicts the schematic of the two steady-state methods for measuring in-plane thermal conductivity along suspended thin films. Figure 9a shows the measurement configuration that was first developed by Volklein and co-workers [76,77], where a metallic strip (Cr/Au film) is deposited on top of the thin film that serves as both an electrical heater and temperature sensor. When a direct current (DC) passes through the heater/sensor, the temperature rise in the heater/sensor is a function of the heating power, thin film thermal conductivity, ambient temperature, thin film thickness $d_f$ and width $L_f$. The in-plane thermal conductivity can then be deduced from the difference in heater/sensor temperature rise of two measurements using two different thin film widths with all other parameters unchanged [77]. The other steady-state method is shown in Figure 9b, another sensor is used to measure the heat sink temperature where the thermal conductivity can then be straightforwardly written as:

$$k_{f,\parallel} = \frac{QL_f}{2d_f\left(T_{f,1} - T_{f,2}\right)} \tag{19}$$

where $Q$ is the power dissipated in the metallic heater per unit length, $L_f/2$ is the distance from the heater to the heat sink, $T_{f,1}$ is the thin film temperature right underneath the heater/sensor, which is assumed the same temperature as heater/sensor. $T_{f,2}$ is the temperature of the thin film edge in contact with the substrate.

For measuring the thermal conductivity of an electrically conductive material or a semi-conducting material, an additional electrical insulation layer is needed between the electrical heater/sensor and thin film for both methods, which significantly complicated the data analysis. In order to ensure one-dimensional heat conduction inside thin film, parasitic heat losses must be minimized, which include heat conduction loss along the length direction of the heater/sensor, convection and radiation losses to the ambient. Heat conduction along the heater/sensor can be minimized through advanced microfabrication to minimize its cross-sectional area. To minimize heat loss to the ambient, the experimental measurement is usually carried out under vacuum. Usually a small temperature difference is used to minimize the radiation heat loss while coating the surface with low-emissivity material could be another option. Nevertheless, the most effective way to deal with radiation heat loss would be using a transient heating (e.g. use alternating current) and temperature sensing technique such as the 3ω method.

## 3.2 Transient methods

### 3.2.1 3ω method

The 3ω method is widely used to measure thermal properties of both bulk materials and thin films after it was first introduced in 1990 by Cahill and coworkers [78,79]. Figure 10 shows a typical schematic of the 3ω measurement. The thin film of interest is grown or deposited on a substrate (e.g. silicon, sapphire [80]). A metallic strip (e.g. aluminum, gold, platinum) is deposited on top of a substrate or the film-on-substrate stack. Dimensions of the metallic strip are usually half-width $a$ = 10-50 μm and length $L$ = 1000-10000 μm which is treated as infinitely long in the mathematical model. The metallic strip serves as both an electrical heater and a temperature sensor, as shown in Figure 10. An alternating current (AC) at frequency ω passes through the heater/sensor, which is expressed as:

$$I(t) = I_0 cos(\omega t) \tag{20}$$

where $I_0$ is the current amplitude, which results in Joule heating of the resistive heater/sensor at 2ω frequency because of its electrical resistance. Such a 2ω heating leads to temperature change of the heater/sensor also at 2ω frequency:

$$\Delta T(t) = \Delta T_0 cos(2\omega t + \varphi) \tag{21}$$

where $\Delta T_0$ is temperature change amplitude and $\varphi$ is phase. The temperature change perturbs the heater/sensor's electrical resistance at 2ω:

$$R_e(t) = R_{e,0}(1 + \alpha_R \Delta T) = R_{e,0}[1 + \alpha_R \Delta T_0 cos(2\omega t + \varphi)] \tag{22}$$

where $\alpha_R$ is the temperature-coefficient of resistance of the heater/sensor. $R_{\mathrm{e},0}$ is heater/sensor's electrical resistance at the initial state. When multiplied by the 1ω driving current, a small voltage signal across the heater/sensor at 3ω frequency can be detected [81]:

$$\begin{aligned} V(t) &= I(t)R(t) \\ &= R_{e,0} I_0 cos(\omega t) + \frac{1}{2} R_{e,0} I_0 \alpha_R \Delta T_0 cos(\omega t + \varphi) \\ &\quad + \frac{1}{2} R_0 I_0 \alpha_R \Delta T_0 cos(3\omega t + \varphi) \end{aligned} \tag{23}$$

This change in voltage at 3ω frequency (i.e. the third term in Eq. (23)) has the information about thermal transport within the sample [82]. However, since the 3ω voltage signal (amplitude $R_0 I_0 \alpha_R \Delta T_0/2$) is very weak and usually about three orders of magnitude smaller than the amplitude of the applied 1ω voltage, a lock-in amplifier is usually employed for implementing such measurement technique.

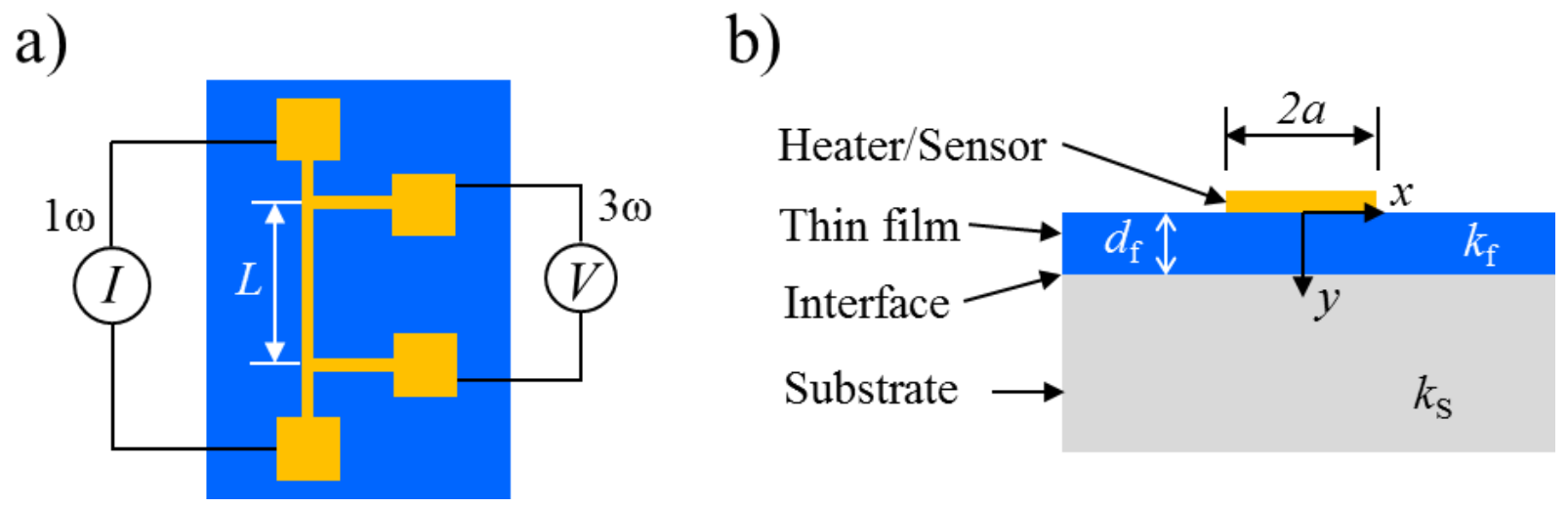

*Figure 10. a) Top view and b) cross-sectional view of a typical 3ω method for thermal characterization of thin films. A metallic strip (i.e. heater/sensor) with width 2a is deposited on top of the thin film, which functions as both heater and temperature sensor. An alternating current at angular frequency 1ω heats the heater/sensor at 2ω frequency. The change in temperature of the heater/sensor causes a change in the resistance which in turn produces a small change in the voltage at 3ω frequency. An electrical insulation layer (not shown in the figure) is required between the thin film and the heater/sensor when measuring electrical conducting thin films.*

When measuring the thermal conductivity of an electrically conductive material or a semi-conducting material, an additional electrical insulation layer is needed between the electrical heater/sensor and thin film. Depending on the width of the heater, both the cross-plane and the in-plane thermal conductivity of thin films can be measured using the 3ω method. Approximate analytical expressions are usually employed to determine the cross-plane and the in-plane thermal conductivity. Borca-Tasciuc *et al.* presented a general solution for heat conduction across a multilayer-film-on-substrate system [13]. Dames also described the general case of thermal and electrical transfer functions framework [82].

For the simplest case that a metallic heater/sensor is deposited on an isotropic substrate without a thin film, the heater/sensor can be approximated as a line source if the thermal penetration depth $L_p = \sqrt{\alpha_S/2\omega}$ is much larger than the heater/sensor half width $a$. By choosing an appropriate heating frequency of the heating current, thermal penetration can be localized within the substrate. The temperature rise of the heater/sensor can then be approximated as [13]:

$$\begin{aligned}\Delta T_S &= \frac{p}{\pi L k_S}\left[0.5\ln\left(\frac{\alpha_S}{a^2}\right) - 0.5\ln(\omega) + \eta\right] - i\left(\frac{p}{4Lk_S}\right) \\ &= \frac{p}{\pi L k_S} f_{\text{linear}}(\ln\omega)\end{aligned} \tag{24}$$

where subscript $S$ is associated with substrate, $\eta$ is a constant, $i$ is $\sqrt{-1}$, $k$ is thermal conductivity, $p/L$ is peak electrical power per unit length, and $f_{\text{linear}}$ is a linear function of $\ln\omega$. It is clear that the isotropic thermal conductivity of substrate $k_S$ can be determined from the slope of the real part of the temperature amplitude as a linear function of the logarithm frequency $\ln(\omega)$ (i.e. the "slope method"), according to Eq. (24).

With a film on substrate, one needs to estimate the temperature drop across the thin film to find the cross-plane thermal conductivity $k_{f,\perp}$ (Figure 11a). The temperature at the upper side of the film is usually taken to be equal to the heater/sensor temperature because the contact resistances are typically very small, $10^{-8} - 10^{-7}\text{m}^2\text{K/W}$ [83]. The most common method to determine the temperature at the bottom side of the film is calculated from the experimental heat flux with the substrate thermal conductivity $k_S$, which is usually known or can be measured using the 3ω method as shown in Eq. (24). Assuming one-dimensional heat conduction across the thin film (Figure 10a), the thermal conductivity of the thin film can be easily determined from:

$$\Delta T_{S+f} = \Delta T_S + \frac{pd_f}{2aLk_{f,\perp}} \tag{25}$$

where subscript f denotes thin film properties, subscript S+f denotes thin film on substrate structure. $k_{f,\perp}$ is obtained by fitting the experimentally measured temperature rise data under a variety of heating frequency ω to Eq. (25) (see Figure. 12).

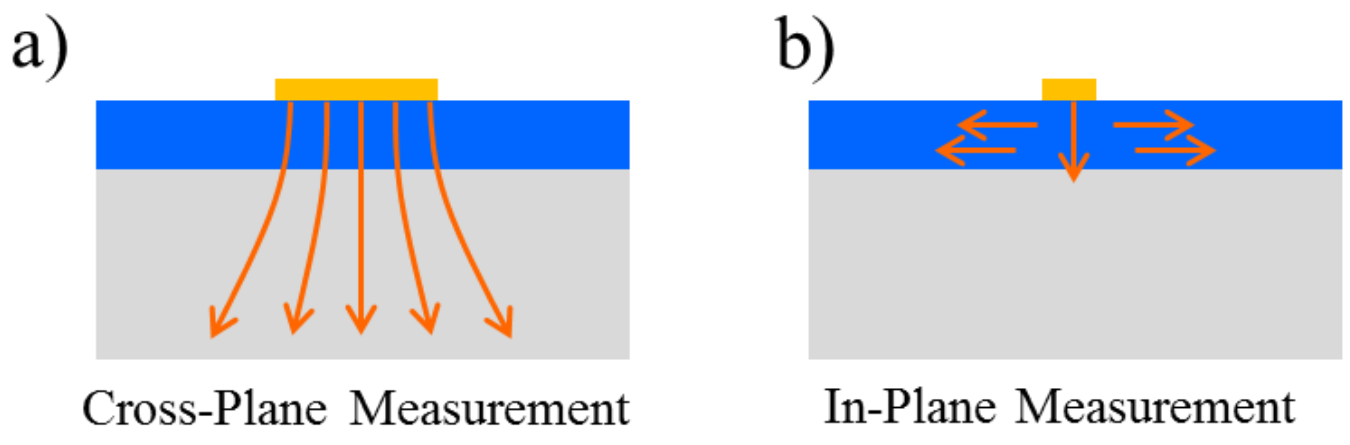

*Figure 11. Schematic of the cross-plane and the in-plane thermal conductivity measurement using the 3ω method. For cross-plane thermal conductivity measurement, the heater width should be relatively large compared to thin film thickness in order to satisfy the assumption that the heat conduction is one-dimensional across the thin film. For the in-plane thermal conductivity measurement, narrower-width heater is used so that the in-plane thermal conductivity can be deduced through heat spreading in the thin film.*

The 3ω method has also been extensively used to measure the in-plane thermal conductivity of thin films. In comparison to the cross-plane thermal conductivity measurement, a much narrower heater is used so that the heat transfer process within the film is sensitive to both in-plane and cross-plane thermal conductivity as shown in Figure 11b. The half width $a$ of the heater should be narrow enough to satisfy [82]:

$$\frac{a}{d_f}\left(\frac{k_{f,\perp}}{k_{f,\parallel}}\right)^{1/2} \leq 0.1 \tag{26}$$

where $k_{f,\perp}$ and $k_{f,\parallel}$ are the cross-plane and in-plane thermal conductivities of the thin film, respectively. $d_f$ is the film thickness. Due to the lateral heat spreading which is sensitive to the in-plane thermal conductivity, a two-dimensional heat transfer model needs to be used for data reduction. The temperature drop across the thin film is obtained as [13]:

$$\Delta T_f = \frac{p}{\pi L}\left(\frac{1}{k_{f,\perp}k_{f,\parallel}}\right)^{1/2}\int_0^{\infty}\frac{sin^2\lambda}{\lambda^3}tanh\left[\lambda\left(\frac{d_f}{a}\right)\left(\frac{k_{f,\parallel}}{k_{f,\perp}}\right)^{1/2}\right]d\lambda \quad (27)$$

Eq. (27) gives the temperature drop of thin film normalized to the value for purely one-dimensional heat conduction through the film, as a function of the in-plane and cross-plane thermal conductivities and the heater/sensor half width $a$. In practice, $k_{f,\perp}$ is usually measured first with a heater/sensor with much greater width that is only sensitive to cross-plane thermal conductivity. $k_{f,\parallel}$ is then measured with a much smaller heater/sensor width (see Figure 12).

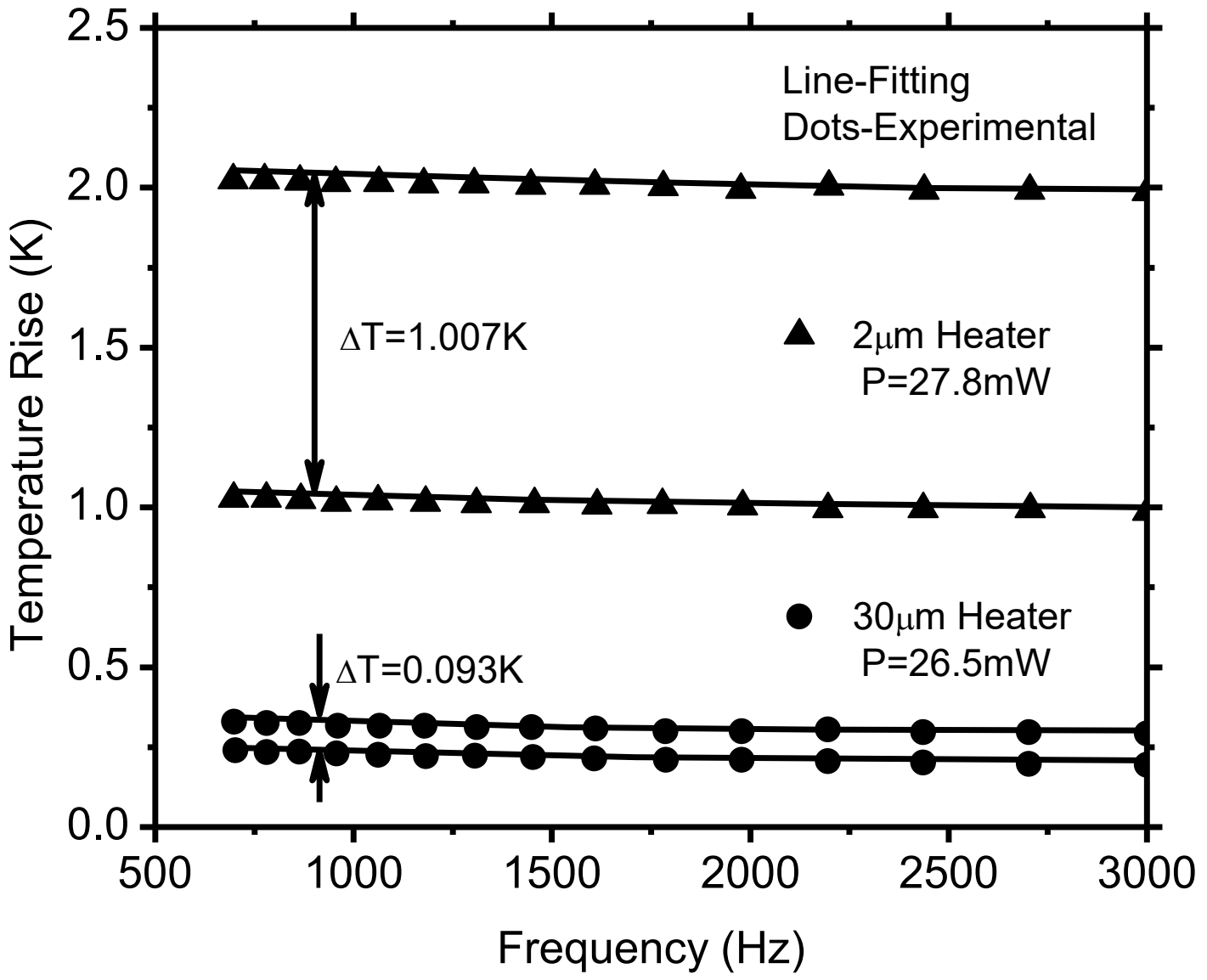

*Figure 12. Experimental temperature rise (dots) measured by 30 and 2μm width heater/sensor deposited on the Si/Ge quantum dot superlattice and the reference samples without the superlattice film of interest. The experimental signal is compared to the predictions for the temperature rise of the heater/sensor*

*calculated based on Eqs. (24), (25) and (27) for the fitted values of in-plane and cross-plane thermal conductivity of the film. The figure is reconstructed from Ref. [*13*]*

The above-mentioned 3ω method has been limited to samples with thermal conductivity tensors that are either isotropic or have their principle axes aligned with the Cartesian coordinate system defined by the heater line and the sample surface. Recently, Mishra *et al.* introduced a 3ω method that can measure an arbitrarily aligned anisotropic thermal conductivity tensor [84]. An anisotropic thermal conductivity tensor with finite off-diagonal terms is considered. An exact closed-form solution has been found and verified numerically. The authors found that the common "slope method" yields the determinant of the thermal conductivity tensor, which is invariant upon rotation about the heater's axis. Following the analytic result, an experimental scheme is proposed to isolate the thermal conductivity tensor elements. By using four heater lines, this method can measure all 6 unknown tensor elements in the 3 dimensions.

A significant advantage of the 3ω method over the conventional steady-state methods is that the error due to radiation heat loss is greatly reduced. Errors due to thermal radiation is shown to scale with a characteristic length of the experimental geometry. The calculated error of 3ω measurement due to radiation is less than 2% even at a high temperature of 1000 K [85]. 3ω method can be used for measuring dielectric, semiconducting and electrically conducting thin films. For electrically conducting and semiconducting materials, samples need to be electrically isolated from the metallic heater/sensor with additional insulating layer [80,86], which introduces extra thermal resistance and inevitably reduces both sensitivity and measurement accuracy. Another challenge is that the 3ω method involves microfabrication for the metallic heater/sensor. Optical heating and sensing method (e.g. transient thermoreflectance technique), on the other hand, usually requires minimal sample preparation.

### *3.2.2 Transient thermoreflectance technique*

The transient thermoreflectance technique is a non-contact optical heating and sensing method to measure thermal properties (thermal conductivity, heat capacity, and interfacial thermal conductance) of both bulk and thin film materials. Samples are usually coated with a metal thin film (e.g. Aluminum or Tungsten), referred to as metallic transducer layer, whose reflectance changes with the temperature rise at the laser wavelength. This allows to detect the thermal response by monitoring the reflectance change. Figure 13 shows schematic diagrams of the sample configuration for a thin film and a bulk material being measured using concentrically focused pump and probe beams. The thermoreflectance technique was primarily developed in the 1970s and 1980s when continuous wave (CW) light sources are used to heat the sample [87,88]. With the advancement of pico- and femtosecond pulsed laser after 1980, this technique has been widely used for studying non-equilibrium electron-phonon interaction [89,90], detecting coherent phonon transport [91–93] and thermal transport across interfaces [94–96] . This technique has recently been further developed over the last few years for measuring anisotropic thermal conductivity of thin films [97–99] and probing spectral phonon transport [100–103].

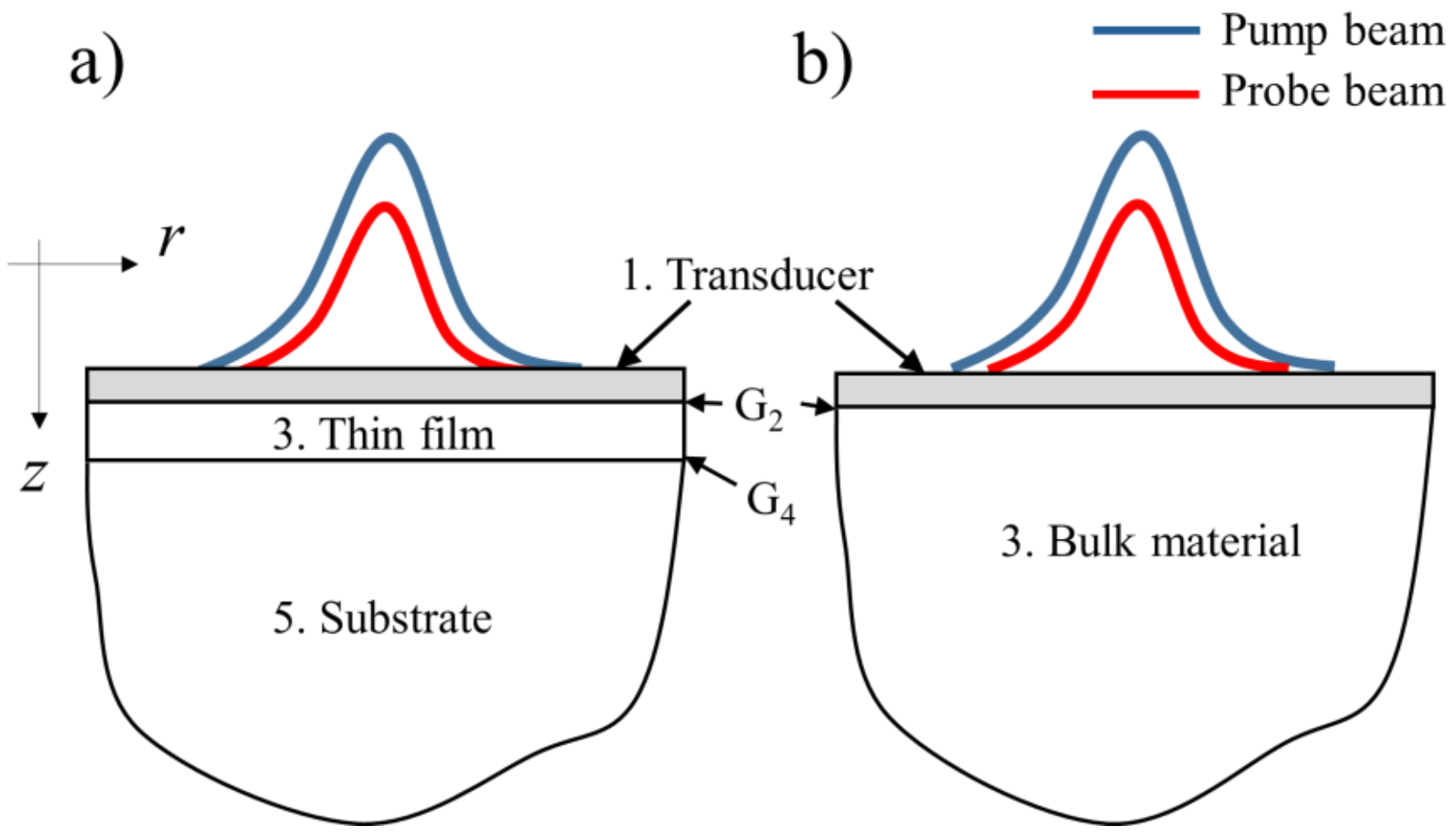

*Figure 13. Typical sample configuration used for measuring thermal properties of a) thin film on a substrate and b) bulk material, using the transient thermoreflectance technique with concentric pump and probe beams. The sample is heated by the frequency-modulated pump laser pulse. The change in the temperature-dependent reflectivity of the metal transducer is measured by the probe laser pulse that is delayed in a picosecond to nanosecond time scale. In the Figure,* $G_2$ *is the interfacial thermal conductance between transducer and the sample.* $G_4$ *is the interfacial thermal conductance between thin film and substrate.*

The transient thermoreflectance technique can be implemented as both time-domain thermoreflectance (TDTR) method [104,105] and frequency-domain thermoreflectance (FDTR) method [98,106]. The TDTR method measures the thermoreflectance response as a function of the time delay between the arrival of the probe and the pump pulses at the sample surface. Figure 14a shows a typical experimental system. A Ti-sapphire oscillator is used as the light source with wavelength centered at around 800 nm and a repetition rate of 80 MHz. The laser output is split into pump beam for heating and probe beam for sensing. Before focused onto the sample, the pump beam is modulated by an acoustic-optic modulator (AOM) or electro-optic modulator (EOM) at a frequency from a few kHz to a few MHz. The probe beam passes through a mechanical delay stage such that the temperature responses are detected with a delay time (usually a few picoseconds to a few nanoseconds) after the sample is heated by the pump pulse. The signal from thermoreflectance change is then extracted by the lock-in amplifier. Spatially separating the pump and the probe, or spectrally screening the pump beam with a filter ( the two-color [107] and two-tint method [108]) can avoid the scattered light from modulated pump beam entering the photodetector.

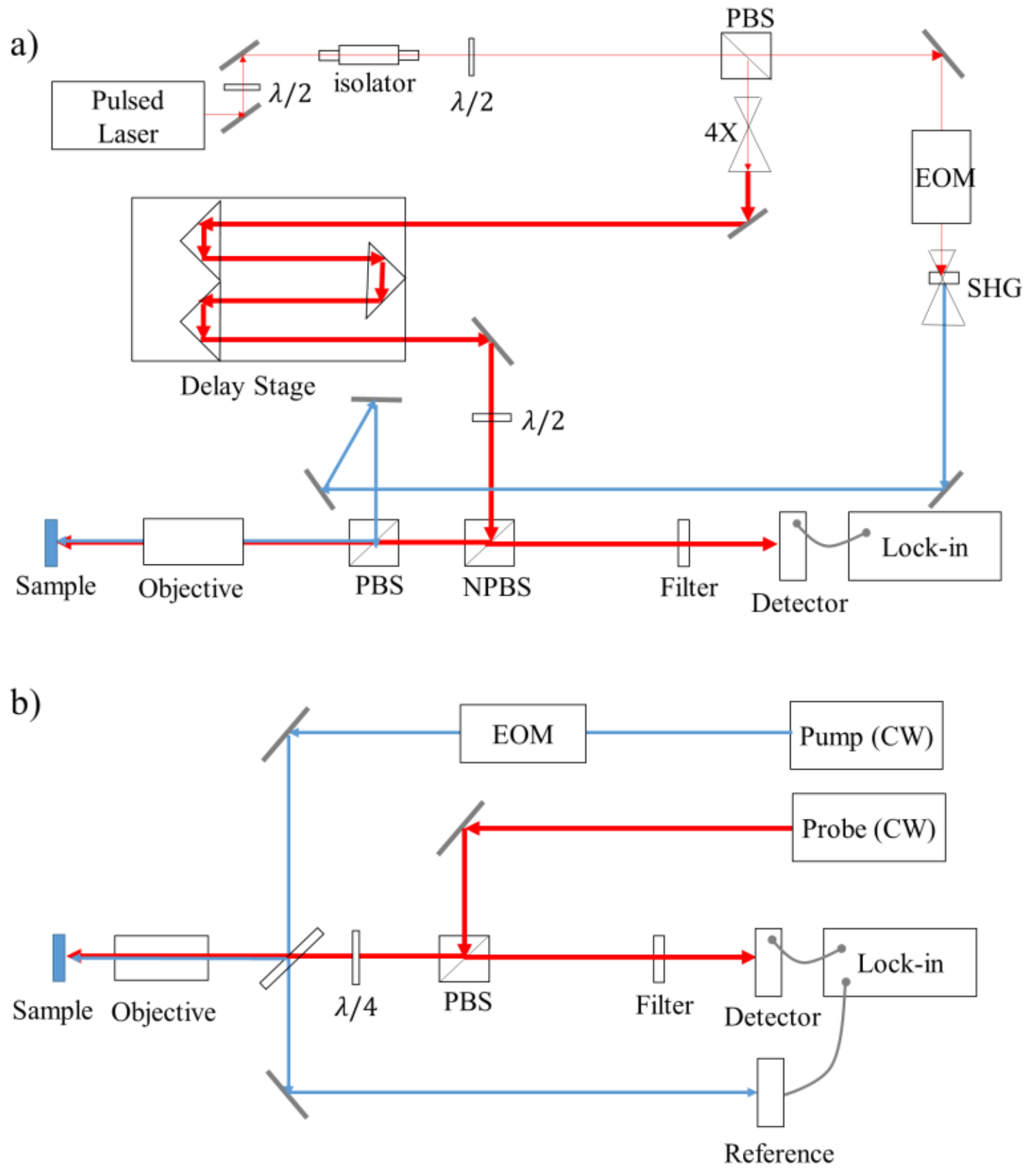

*Figure 14. a) Schematic of a two-color TDTR system. After the first polarized beam splitter (PBS), the probe beam passes through a beam expander to minimize the spot size change over length of the delay stage, and it is then directed onto the sample through an objective lens at normal incidence. The pump beam passes through a second harmonic generation (SHG) crystal for frequency doubling and then the electro-optic modulator (EOM) for modulation before being directed through the same objective onto the sample, coaxial with the probe beam. This system can also be used for FDTR measurement by fixing the delay stage and varying the modulation frequency. b) Schematic of a FDTR system based on two CW lasers with different wavelengths. The pump beam is modulated by EOM to create a periodic heating and the probe beam measures the thermoreflectance change. This FDTR system is modified from Ref.* [107], *Copyright Begell House Digital Library.*

The other group of thermoreflectance technique is FDTR, where the thermoreflectance change is measured as a function of the modulation frequency of the pump beam. Therefore, FDTR can be easily implemented using the same TDTR system (Figure 14a) by fixing the delay stage at a certain position and varying the modulation frequency. FDTR system, however, can avoid the complexity of beam walk-off and divergence associated with the mechanical delay stage because the probe delay is fixed. FDTR system can also be implemented using the less expensive continuous wave (CW) lasers as shown in Figure 14b, which achieves similar accuracy to TDTR measurement for thermal conductivity of many thin-film materials [98,105,106,109]. Similar to TDTR, the pump beam of CW based FDTR (CW-FDTR) is modulated by EOM and creates a time dependent temperature gradient. However, the heating process by pump incidence is continuous in CW-FDTR. The probe beam is directly focused on the sample without passing through a mechanical delay stage. The thermoreflectance change embedded in the reflected probe beam is also extracted using a photodiode and a lock-in amplifier.

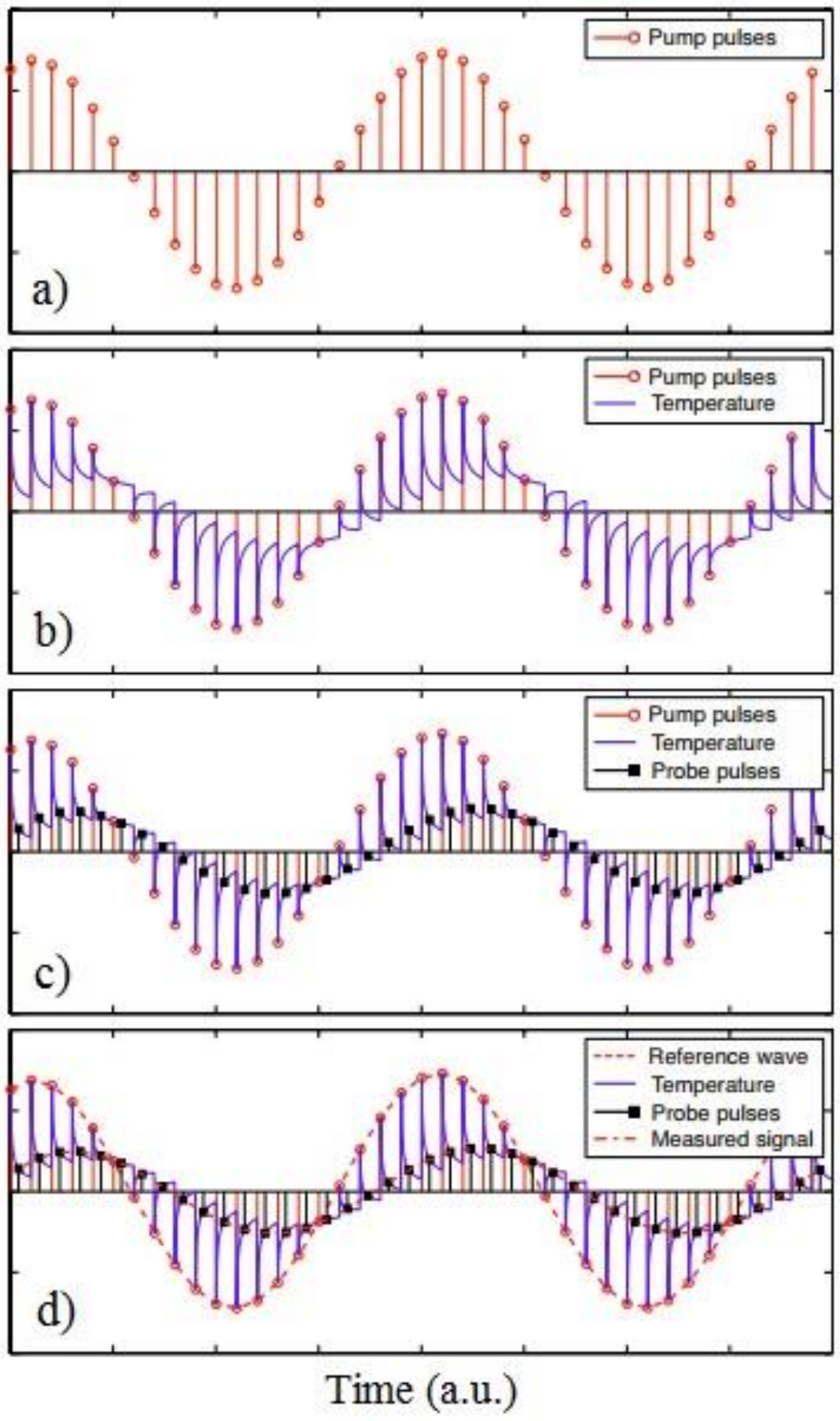

*Figure 15. An illustration of the TDTR detection scheme: a) The pump pulse input to the sample is modulated by the electro-optic modulator or acoustic-optic modulator. b) Sample surface temperature response to the pump pulse input. c) The probe pulses arrive at the sample with a delayed time and are reflected back to a detector with an intensity proportional to the surface temperature. d) The fundamental harmonic components of the reference wave and measured probe wave. The amplitude and phase difference between these two waves are picked up by a lock-in amplifier. From Ref. [99], Copyright 2008, AIP Publishing.*

An illustration of the TDTR data acquisition is given in Figure 15. The pump pulses modulated at frequency $\omega_0$ (Figure 15a) heat the sample periodically. The oscillating temperature response of the sample (Figure 15b) is then detected by the probe beam arriving with a delayed time $\tau_0$. In this case, the thermoreflectance response $Z$ at modulation frequency $\omega_0$ is expressed as an accumulation of the unit impulse responses $h(t)$ in time domain:

$$Z(\omega_0, \tau_d) = V_{in} + iV_{out} = \beta \frac{2\pi}{\omega_s} \sum_{n=0}^{\infty} h\left(n\frac{2\pi}{\omega_s} + \tau_d\right) e^{-i\omega_0(nT_0+\tau_d)} \quad (28)$$

where $V_{in}$ and $V_{out}$ are the real and imaginary part of the response usually referred to as in-phase and out-of-phase signal respectively, $\tau_d$ is the delay time, $\frac{2\pi}{\omega_s}$ is the time between two successive pulses at laser repetition frequency $\omega_s$ and $\beta$ is a constant coefficient determined by:

$$\beta = G_{det} P_1 (1 - R_1) \left(\frac{dR_2}{dT}\right) P_2 R_2 \quad (29)$$

where $G_{det}$ is the gain of photodetect, $P_1$ and $P_2$ are the power of pump and probe beams respectively, $R_1$ and $R_2$ are the reflectivity at the wavelengths of pump and probe beams respectively, and $\frac{dR_2}{dT}$ is the thermoreflectance coefficient of the transducer at probe wavelength. Identically, the thermoreflectance response $Z$ can be expressed in the frequency domain:

$$Z(\omega_0, \tau_d) = V_{in} + iV_{out} = \beta \sum_{l=-\infty}^{\infty} H(\omega_0 + l\omega_s) e^{il\omega_s \tau_d} \quad (30)$$

where $H(\omega_0 + l\omega_s)$ is the thermoreflectance response of the sample heated by a continuous Gaussian beam modulated at frequency $(\omega_0 + l\omega_s)$. The single frequency response $H(\omega)$ is determined by thermal properties including thermal conductivity $k$, heat capacity $C$ of each layer, and interfacial thermal conductance $G$ between different layers.

Here we show an outline for the derivation of the single frequency response function $H(\omega)$, while the detailed derivation of the heat transfer model can be found in Refs. [94,105,107]. The heat conduction equation in cylindrical coordinates is written as:

$$C \frac{\partial T}{\partial t} = \frac{k_r}{r} \frac{\partial}{\partial r}\left(r \frac{\partial T}{\partial r}\right) + k_z \frac{\partial^2 T}{\partial z^2} \quad (31)$$

where $k_r$ and $k_z$ are thermal conductivity in the in-plane and cross-plane directions. (see Figure 16 for definition of in-plane and cross-plane directions).

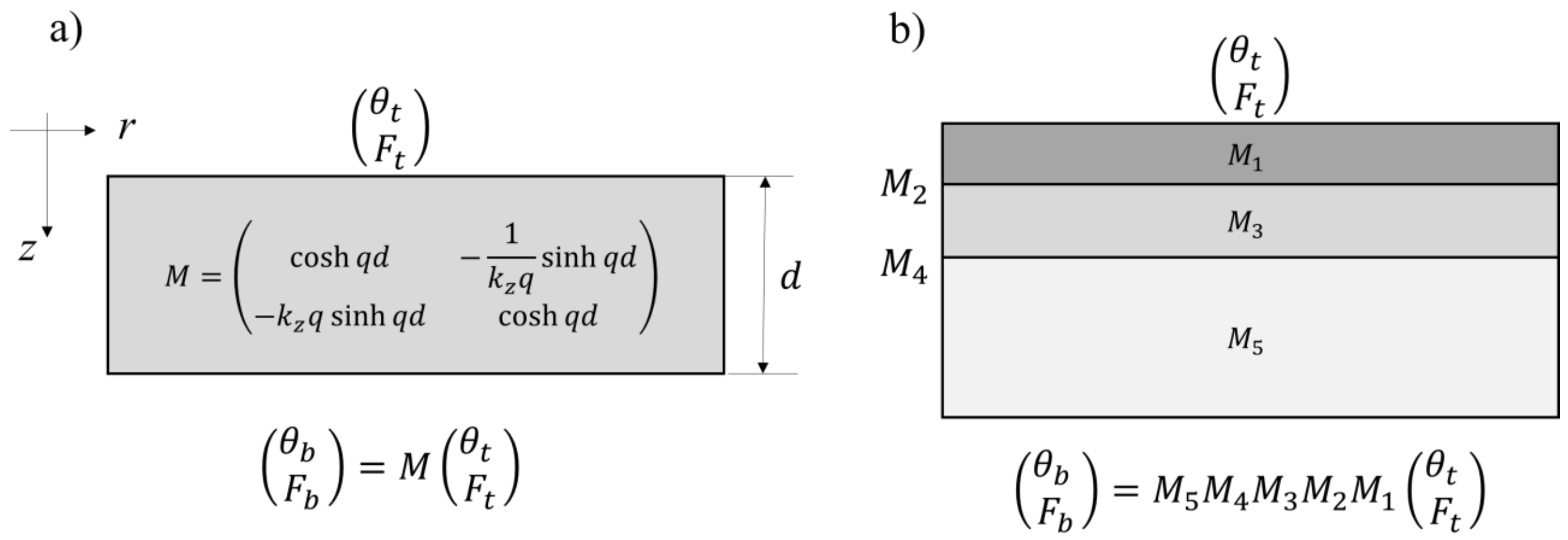

*Figure 16. Schematic of heat transfer model for deriving single frequency response in cylindrical coordinate. The r denotes in-plane direction and z denotes cross-plane direction. a) The transfer matrix of a single layer and b) the transfer matrices of multilayers. θ and F denotes temperature and heat flux respectively after Fourier and Hankel transform, and the subscripts t and b denotes top and bottom sides.*

By applying the Fourier transformation to the time variable $t$ and Hankel transform to the radial coordinate $r$, a transfer matrix can be obtained relating the temperature $\theta_t$ and heat flux $F_t$ on the top of the single slab and the temperature $\theta_b$ and heat flux $F_b$ on the bottom side (Figure 16a):

$$\begin{pmatrix}\theta_b\\F_b\end{pmatrix} = \begin{pmatrix}\cosh qd & -\frac{1}{k_z q}\sinh qd\\ -k_z q\sinh qd & \cosh qd\end{pmatrix}\begin{pmatrix}\theta_t\\F_t\end{pmatrix} \qquad (32)$$

where $d$ is the thickness of the slab, and the complex thermal wave number $q = \sqrt{\frac{k_r x^2 + i\omega C}{k_z}}$, where $x$ is the Hankel transform variable. Multiple layers can be handled by multiplying the matrices for individual layers together (Figure 16b):

$$\begin{pmatrix} \theta_b \\ F_b \end{pmatrix} = M_n M_{n-1} \dots M_1 \begin{pmatrix} \theta_t \\ F_t \end{pmatrix} = \begin{pmatrix} A & B \\ C & D \end{pmatrix} \begin{pmatrix} \theta_t \\ F_t \end{pmatrix} \tag{33}$$

For interfaces, the transfer matrix can be obtained by taking the heat capacity as zero and choosing $k_z$ and $d$ such that interfacial thermal conductance $G$ equals to $k_z/d$. The boundary condition is approximately adiabatic in experimental condition, hence $F_b = C\theta_t + DF_t = 0$, the surface temperature can be obtained by:

$$\theta_t = -\frac{D}{C} F_t \tag{34}$$

In experiment, if a Gaussian laser with radius $w_0$ and power $Q$ is used as pump, the heat flux $F_t$ after Hankel transform is $F_t = \frac{Q}{2\pi} \exp(-(w_0^2 x^2)/8)$ , then Eq. (34) becomes

$$\theta_t = -\frac{D}{C} \frac{Q}{2\pi} \exp\left(-\frac{w_0^2 x^2}{8}\right) \tag{35}$$

The reflectivity response is then weighted by the Gaussian distribution of probe beam intensity with radius $w_1$:

$$H(\omega) = \frac{Q}{2\pi} \int_0^\infty x \left(-\frac{D}{C}\right) \exp\left(-\frac{(w_0^2 + w_1^2) x^2}{8}\right) dx \tag{36}$$

We give an example of TDTR signal detected by the lock-in amplifier in Figure 17. Thermal properties including the thermal conductivity, heat capacity and interfacial thermal conductance are encoded in the TDTR signal trace. The in-phase signal $V_{in}$ represents the change of surface temperature. The peak in the $V_{in}$ represents the surface temperature jump right after the pump pulse incidence, and the decaying tail in $V_{in}$ represents the cooling of surface due to the heat dissipation in the sample. The out-of-phase signal can be viewed as the sinusoidal heating of the sample at modulation frequency $\omega_0$ [110]. When processing the experimental data, the ratio between in-phase and out-of-phase signal -$V_{in}/V_{out}$ is fitted with the heat transfer model to extract thermophysical properties. In this fitting process, numerical optimization algorithms (e.g. quasi-newton [111] and simplex minimization [112]) are used to minimize the squared difference between the experimental data and the heat transfer model, until the value change in thermal properties is smaller than the tolerance (e.g. 1%).

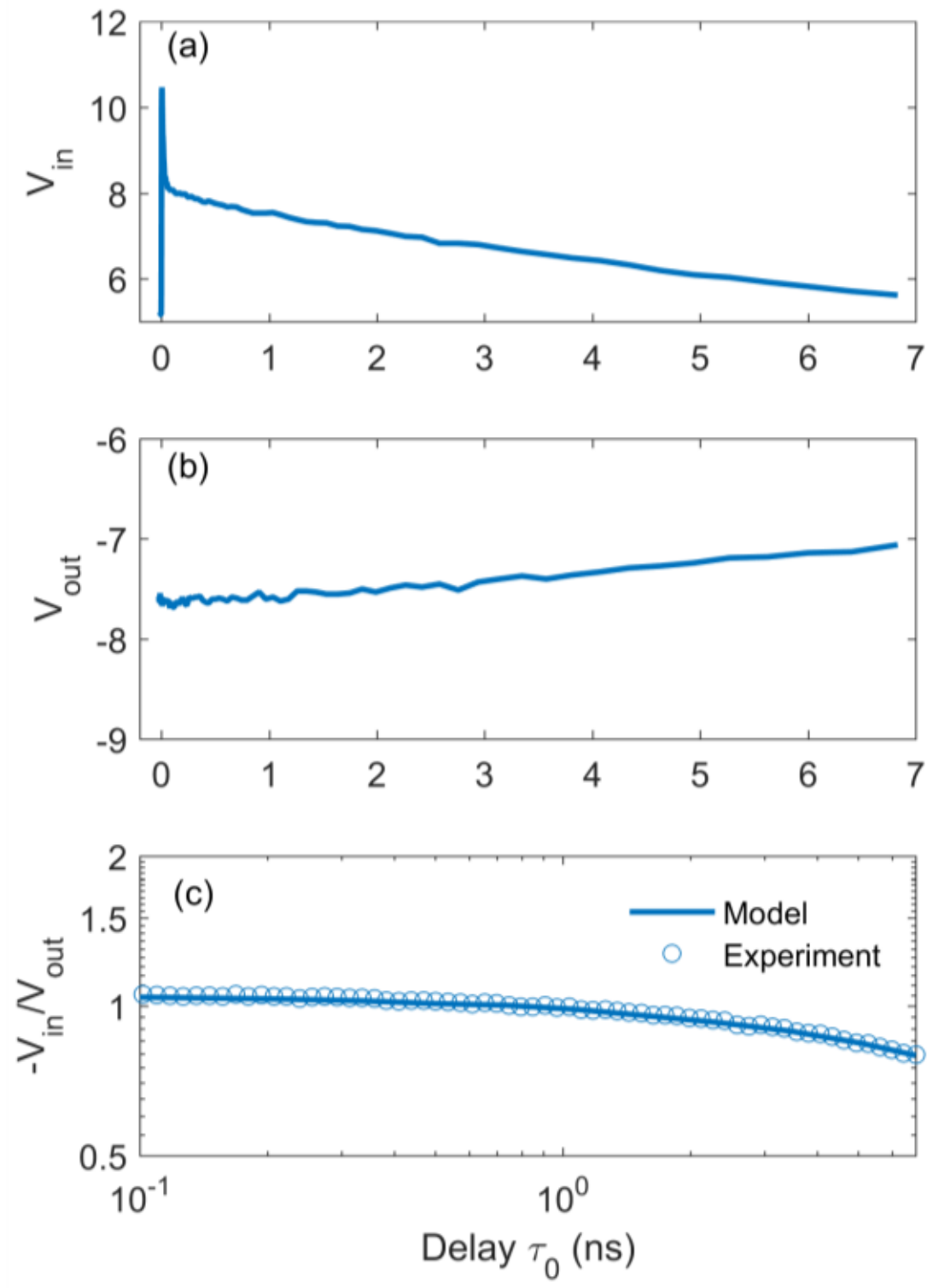

*Figure 17. Measured TDTR temperature responses in terms of (a) in-phase signal (b) out-of-phase signal and (c) the ratio -$V_{in}/V_{out}$ on amorphous $SiO_2$ sample coated with 100 nm Al as metallic transducer. Here a pump spot size of 9.1 μm, a probe spot size of 7.5 μm and a modulation frequency of 1.06 MHz were used. The solid line in (c) is predicted by the heat transfer model with thermal conductivity k=1.4 $W/m{\cdot}K$ and interfacial thermal conductance G=120 $MW/m^2{\cdot}K$.*

In the case of FDTR system based on pulsed laser (Figure 14a), the obtained signal can be fitted with $Z(\omega_0; \tau_d)$ in Eq. (30) where the modulation frequency $\omega_0$ is varied as an independent variable and the delay time $\tau_d$ is fixed. When an FDTR system is implemented using CW lasers, the thermoreflectance signal is instead directly proportional to single frequency response $H(\omega_0)$ [97,113,114]:

$$Z(\omega_0) = \beta H(\omega_0) \tag{37}$$

Figure 18 shows the calculated phase response $\phi = \arctan(V_{out}/V_{in})$ of pulsed and CW-based FDTR measurement of 100 nm Al on sapphire with modulation frequency ranges from 50 kHz to 20 MHz. A clear phase difference is observed between the pulsed FDTR and the CW-based FDTR and attentions must be paid to adopt the correct solution when processing signals from different FDTR systems. Similar to the TDTR system, the least-squared error method can be used to extract thermophysical properties.

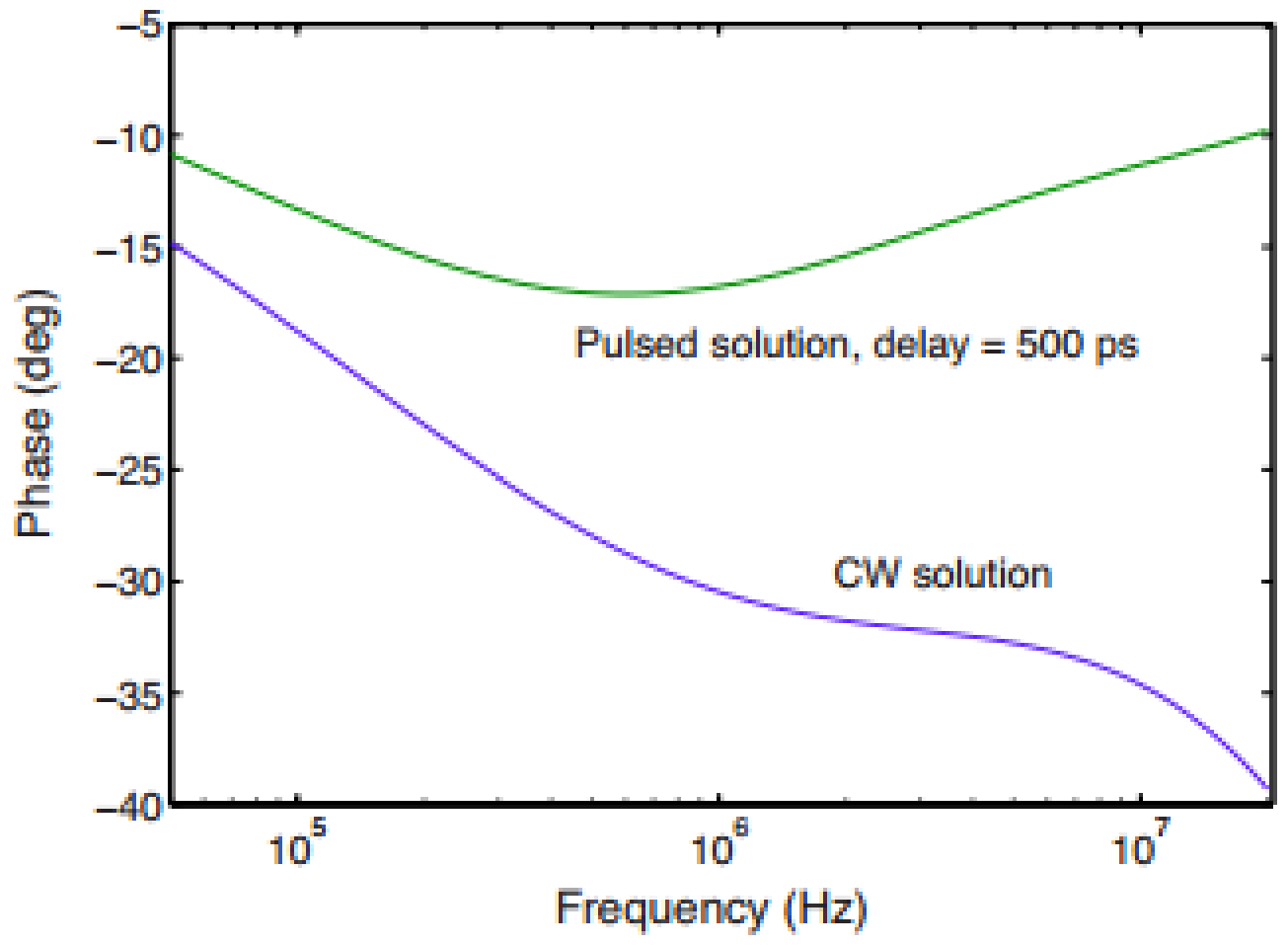

*Figure 18. The calculated phase response for CW and pulsed FDTR of 100 nm Al on sapphire substrate [97]. Copyright 2009, AIP Publishing.*

The challenge of thermoreflectance technique lies in its versatile capability that multiple thermal properties including thermal conductivity, interfacial thermal conductance and heat capacity can be determined depending on the measurement regime. The interfacial thermal conductance can be determined at long delay time (> 2ns) using TDTR [115,116]. A sensitivity parameter $S_p$ is defined to characterize the dependence of signal on different parameters $p$ (interface thermal conductance, heat capacity and thermal conductivity, etc.):

$$S_p = \frac{d\ln(-V_{in}/V_{out})}{d\,(\ln p\,)} \tag{38}$$

The sensitivity parameter describes the scaling law between change in the signal $-V_{in}/V_{out}$ and the change in the parameter $p$. For example, a sensitivity value of 0.4 means that there would be 0.4% change in the signal if the parameter $p$ is changed by 1%. As shown in Figure 19, the TDTR signal is dominantly determined by the interfacial thermal conductance when delay time is longer than 2 ns. TDTR technique is therefore implemented extensively to study the thermal transport mechanisms across interfaces, including the effect of surface chemistry on interfacial thermal conductance across functionalized liquid-solid boundary [95,117,118] and solid-solid interfaces [119], interfacial thermal conductance between metals and dielectrics, [120–124] and interface between low dimensional materials and bulk substrates [125–127]

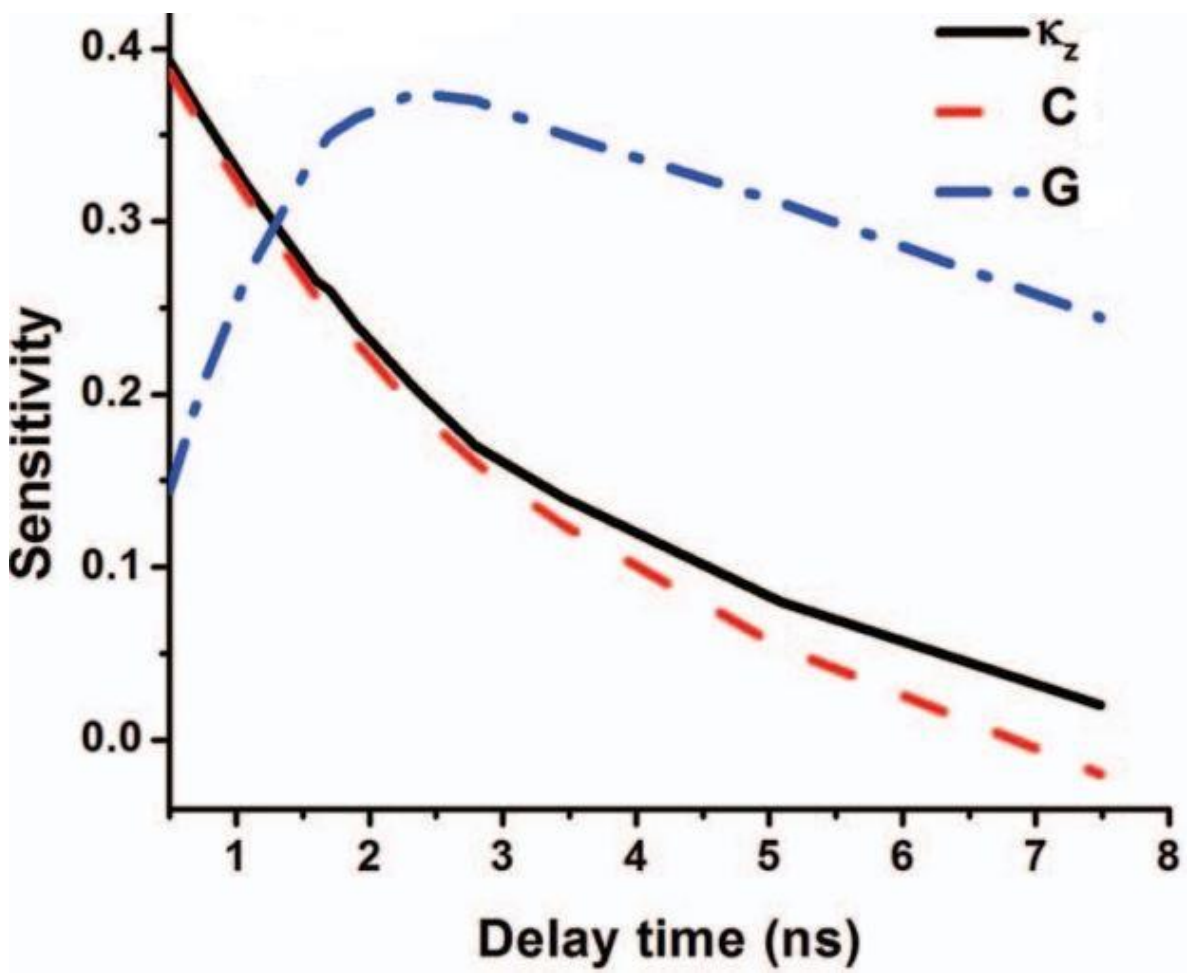

*Figure 19. The sensitivity of $-V_{in}/V_{out}$ signal to the cross-plane thermal conductivity $k_z$ and heat capacity $C$ of silicon, and the interface thermal conductance $G$ between Al and Si at 6.8 MHz. Modified from Ref. [*116*], Copyright 2013, AIP Publishing.*

With interfacial thermal conductance determined, heat capacity and thermal conductivity can then be simultaneously obtained. Liu *et al.* demonstrated that thermal conductivity and heat capacity can be determined simultaneously for both bulk materials and thin films by performing multiple measurements using different combinations of modulation frequencies and spot sizes [116].

Let's consider bulk materials first to introduce the physical picture we based on to determine heat capacity and thermal conductivity simultaneously. Under periodic laser heating, the penetration depth is defined as a characteristic length which describes the depth of temperature gradient penetrates into the sample can be written as $L_{p,\delta} = \sqrt{2k_\delta/\omega_0 C}$ where $\omega_0$ is the modulation frequency, $C$ is volumetric heat capacity and the index of directions $\delta$ corresponds to in-plane ($\delta = r$) and cross-plane ($\delta = z$) respectively. If a very low modulation frequency and small spot size is used to conduct TDTR measurement, the radial heat transfer dominates the thermal dissipation (Figure 20a), and the signal is determined by the geometric average thermal conductivity $\sqrt{k_r k_z}$. This approximation holds true when the radial penetration depth $L_{p,r} \gg \frac{1}{4}\sqrt{w_0^2 + w_1^2}$, where $w_0$ and $w_1$ are $1/e$ radius of pump and probe beam. On the other hand, if we use a large spot size and high modulation frequency, the temperature gradient only penetrates into a very thin layer into the sample and cross-plane thermal transport is dominating (Figure 20b). The thermal response signal is determined by $\sqrt{k_z/C}$. If we assume the material isotropic, we can first determine thermal conductivity using a small beam spot at low frequency, and then measure heat capacity using a large beam spot at high frequency.

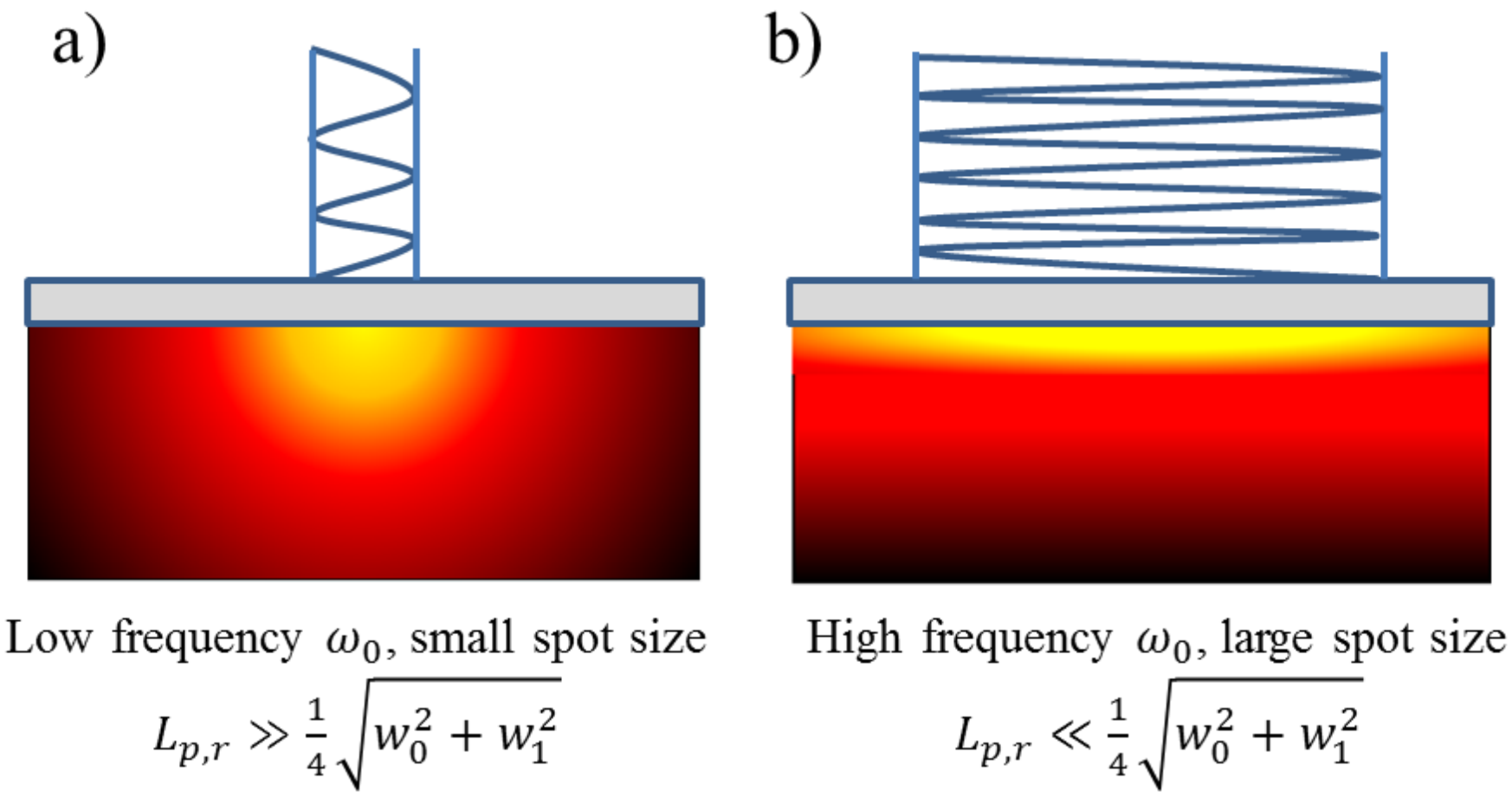

*Figure 20. Schematic of temperature profile with different combination of modulation frequency $\omega_0$ and spot sizes. a) Spherical thermal waves in the sample when heated by a small spot size at low modulation frequency $\omega_0$, and the radial penetration depth $L_{p,r}$ is much longer than $\frac{1}{4}\sqrt{w_0^2 + w_1^2}$ where $w_0$ and $w_1$ are pump and probe beam radius. b) Plane thermal waves in the sample when heated by a large beam spot at high modulation frequency $\omega_0$. The radial penetration depth is much smaller than beam size $\frac{1}{4}\sqrt{w_0^2 + w_1^2}$.*

Measuring thermal properties of thin film is similar to the bulk material with only slight differences. If a large beam spot modulated at high frequency is used to measure thermal properties, the cross-plane heat transfer dominates. Different from bulk samples whose thickness is always much longer than the cross-plane penetration depth, the thickness of a thin film might be comparable or even smaller than the cross-plane penetration depth $L_{p,z}$. We describe the sample "thermally thin" if the penetration depth $L_{p,z}$ is still much larger than film thickness $d$ (Figure 21a). In this case the thermal response is controlled by the thermal resistance $d/k_z$ and heat capacity $C$. At high frequency limit, the penetration depth $L_{p,z}$ would be so small that temperature gradient only penetrates into a limited depth of the layer (referred to as "thermally thick", Figure 21b), the thermal response is analogous to the high frequency limit of the bulk material, only dominated by thermal effusivity $\sqrt{k_z C}$. If the penetration depth is comparable to the thin film thickness, both thermal effusivity $\sqrt{k_z C}$ and diffusivity $k_z/C$ affect the temperature response of the thin film sample. Based on the analysis above, both cross-plane thermal conductivity and heat capacity of thin films can be obtained by applying different modulation frequencies (Figure 22).

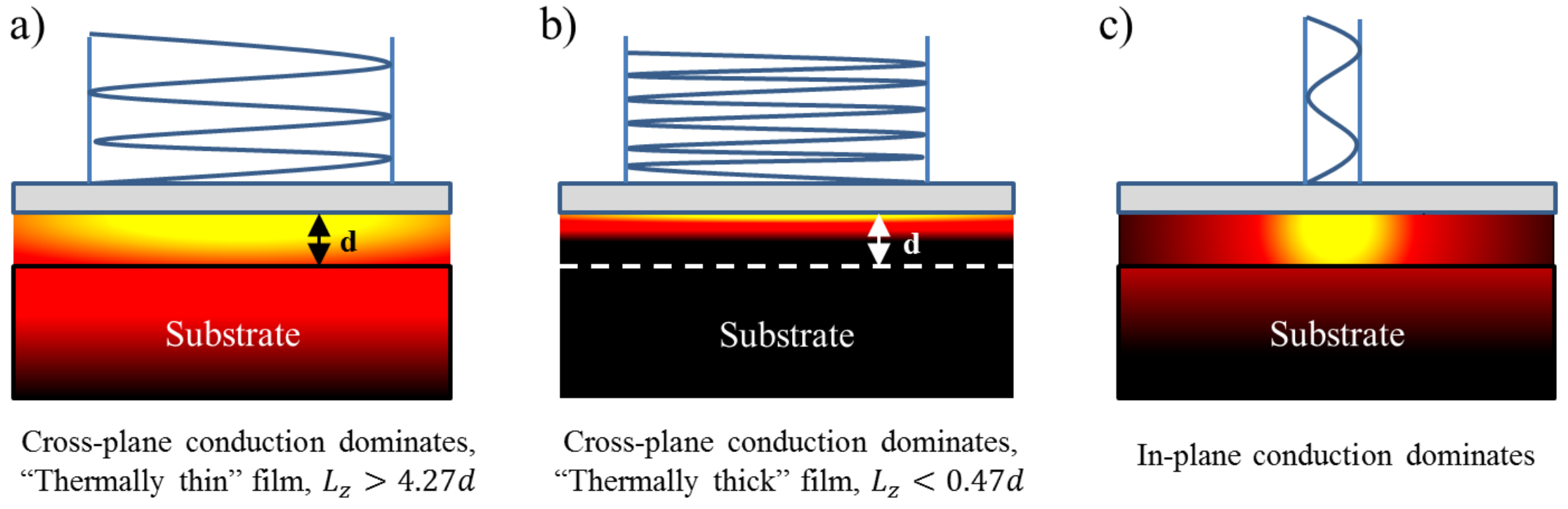

*Figure 21. Schematic of temperature profile when a) a thermally thin film where the cross-plane penetration depth $L_{p,z}$ is much larger than thickness of the film $d$, b) a thermally thick film where the thermal excitation only penetrates into a limited depth into the thin film, and c) in-plane heat transfer dominates with a small beam spot size at low modulation frequency.*

If the beam is tightly focused to the thin film sample at low modulation frequency, the heat transfer is dominated in the in-plane radial direction (Figure 21c). In this case, the in-plane thermal conductivity $k_r$ dominates the thermal response of the material. A low thermal conductivity substrate can be used to further improve the sensitivity to the in-plane thermal conductivity of the thin film [97,99,116].

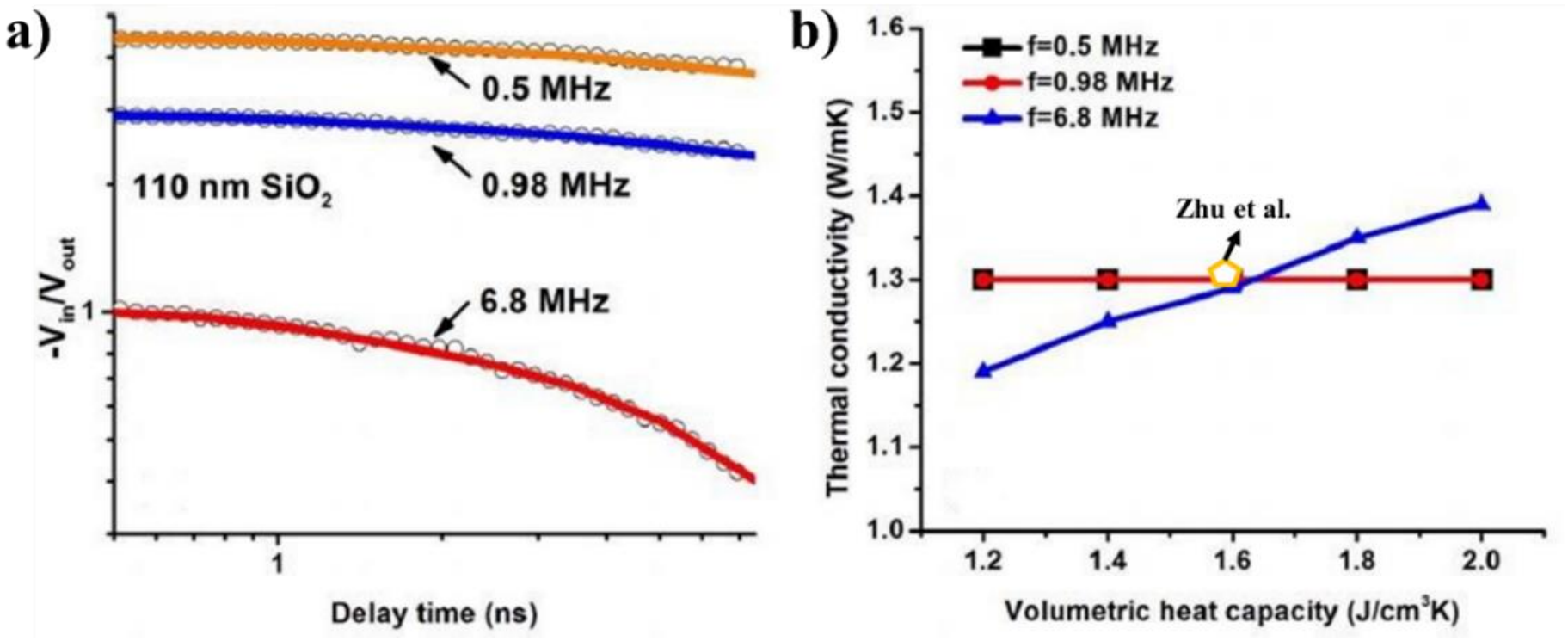

*Figure 22. a) measured TDTR signal for 110 nm $SiO_2$ film at multiple modulation frequencies. b) The thermal conductivity-heat capacity (k, C) pairs yielding best fit to the experiment data are not unique. Simultaneous determination of k and C can be obtained using the k-C diagram. Data is obtained using a 110 nm $SiO_2$, compared with data measured by J. Zhu et al. [*105*]. Modified from Ref. [*116*], Copyright 2013, AIP Publishing.*

This method of measuring in-plane thermal conductivity using a small spot size, however, assumes cylindrical symmetry of thermal conductivity in the in-plane direction [94]. For anisotropic thin films lacking in-plane symmetry, the thermal conductivity tensor can be extracted by offsetting the pump beam away from the probe beam [128]. The schematic of implementing beam-offset TDTR on measuring in-plane thermal conductivity tensor of $\alpha$-$SiO_2$ is shown in Figure 23. Instead of measuring the ratio between in-phase and out-of-phase signal $-V_{in}/V_{out}$, beam-offset TDTR measures the full-width half maximum (FWHM) of the out-of-phase signal $V_{out}$, as the pump beam is detuned from the probe beam (Figure 23b). By sweeping the beam in different directions parallel to the thin film plane, beam-offset TDTR directly detects the in-plane penetration length by monitoring the two-dimensional temperature profile in the in-plane direction, thus makes it possible to extract anisotropic in-plane thermal conductivity. For example, the thermal conductivity $k_c$ along c-axis of the $SiO_2$ can be measured by offsetting the pump beam parallel to the c-axis (Figure 23a), and the out-of-phase signal is then recorded as a function of beam offset distance $y_0$, shown as the open circles in Figure 23b. This step maps a Gaussian shaped out-of-phase signal, and the experimental FWHM is extracted from the Gaussian profile of $V_{out}$. Then the FWHM can be calculated as a function of thermal conductivity $k_a$ shown as the curves in Figure 23c. With the experimentally measured FWHM, the thermal conductivity $k_c$ can be extracted from the FWHM-$k_c$ curve. The FWHM perpendicular to the c-axis is also plotted in Figure 23d, nearly independent of the thermal conductivity $k_c$ along c-axis, ensuring that the in-plane thermal conductivity along different directions can be extracted independently by offsetting the pump beam in different directions. Similarly, the thermal conductivity $k_a$ perpendicular to the c-axis can also be extracted by offsetting the pump beam along $x$-direction (Figure 23c). The beam offset technique allows the TDTR technique to extract thermal conductivity tensor for materials lacking in-plane symmetry. This beam-offset technique has also been extended to time-resolved magneto-optic Kerr effect (TR-MOKE) based pump-probe measurement [129], and has been applied to measure thermal conductivity tensor of two-dimensional materials like $MoS_2$ [129] and black phosphorous [130,131].

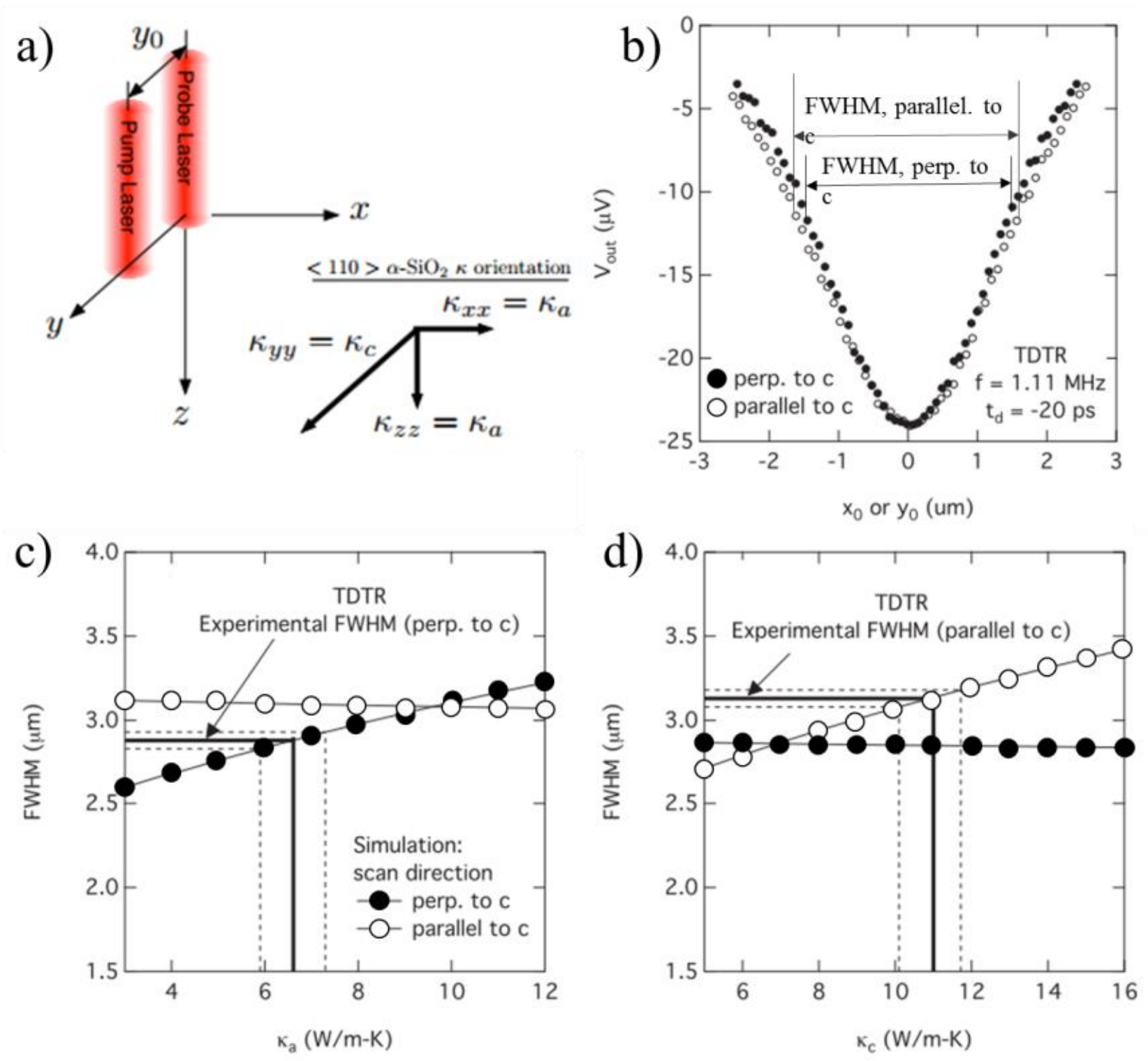

*Figure 23. a) Schematic of beam-offset TDTR and the alignment of the <110> $SiO_2$ principal thermal conductivity axes with the experimental coordinate, b) The experimental beam-offset data of out-of-phase signal, c) comparison of experimental and theoretical FWHM to extract thermal conductivity $k_a$ perpendicular to c-axis and d) comparison of experimental and theoretical FWHM to extract thermal conductivity $k_c$ parallel to c-axis. Modified from [128], Copyright 2014, AIP Publishing LLC.*

Transient thermoreflectance method has been widely applied to explore thermal properties of novel materials that ranging from low thermal conductivity of ≤ 0.1 W/m·K of hybrid materials [132,133] and fullerene derivatives [134,135], to very high thermal conductivity such as graphene [136]. In addition, the capability of determining multiple thermophysical properties simultaneously makes transient thermoreflectance a versatile technique to characterize thermal properties of a wide range of materials including diamond [137,138], pure and doped Si films [115], disordered layered crystals [139], and superlattices [104,140].

## 4. Summary

This work reviews the measurement techniques for thermal conductivity and interfacial contact resistance of both bulk materials and thin films. For thermal characterization of bulk material, the steady-state absolute method, comparative technique, laser flash diffusivity method, and transient plane source (TPS) method are most used. For thin film measurement, the $3\omega$ method and transient thermoreflectance technique are employed widely. Figure 24 gives a summary of sample size and measurement time scale for different thermal conductivity measurement techniques. The absolute technique and comparative technique measure large-size sample but need longest time for data acquisition, while transient thermoreflectance technique is capable of measuring thinnest film sample in a quickest way. In general, it is a very challenging task to determine material thermal conductivity and interface contact resistance with less than 5% error. Selecting a specific measurement technique to characterize thermal properties need to be based on: 1) knowledge on the sample whose thermophysical properties is to be determined, including the sample geometry and size, surface roughness and preparation method; 2) understanding of fundamentals and procedures of the testing technique and equipment, for example, some techniques are limited to samples with specific geometries and some are limited to specific range of thermophysical properties; 3) understanding of the potential error sources which might affect the final results, for example, the convection and radiation heat losses.

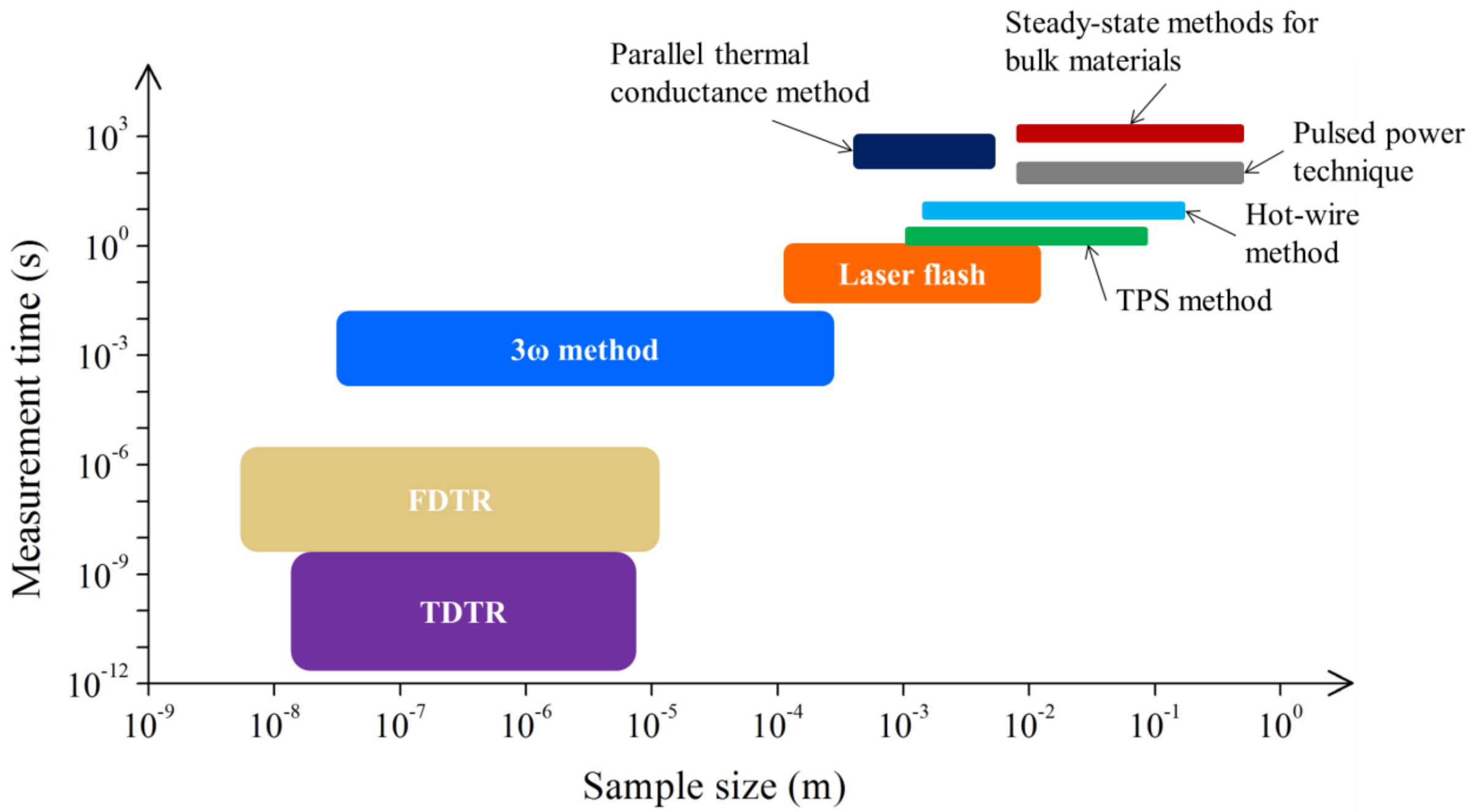

*Figure 24. Sample size and measurement time for different thermal conductivity measurement techniques. The sample size is the length of temperature gradient inside the sample during the measurement. The measurement time is the temperature variation time (continuous or pulsed heating source) or temperature variation period (continuous periodic heating source) during the test. The state-state methods for bulk materials measure large-size sample but need longest time for data acquisition, while transient thermoreflectance technique (TDTR and FDTR) is capable of measuring thinnest film sample in a quickest way. Note that the steady-state methods for bulk materials including absolute technique, comparative technique and radial heat flow method.*